\documentclass[a4paper,11pt]{article}
\pdfoutput=1
\usepackage{jcappub} 
\usepackage{slashed}
\allowdisplaybreaks
\title{\boldmath Scale-dependence  in  DHOST inflation}

\author[a]{Philippe Brax,}
\author[b]{Andrei Lazanu}

\affiliation[a]{Institut de Physique Th\'eorique, Universit\'e  Paris-Saclay, CEA, CNRS, F-91191 Gif-sur-Yvette Cedex, France}
\affiliation[b]{Laboratoire de Physique de l’Ecole normale sup\'erieure, ENS, Universit\'e PSL, CNRS, Sorbonne Universit\'e, Universit\'e de Paris, F-75005 Paris, France}

\abstract{We study the inflationary consequences of Degenerate Higher Order Scalar Tensor (DHOST) theories in a de Sitter background. We  perturb the de Sitter background by  operators breaking either the degeneracy condition, i.e scordatura DHOST, or the shift symmetry in the scalar field. We first consider derivative  scordatura and  find that in all cases the power spectra of curvature perturbations are scale-invariant. We then investigate small perturbations by an axion-like potential, and show that in this scenario the power spectrum becomes scale-dependent. The modifications to the spectral index and its first two derivatives are compatible with the latest inflationary constraints. Moreover the tensor to scalar ratio and the non-Gaussianities of these models  could be within reach of future experiments. }

\begin{document}
\maketitle
\flushbottom

\section{Introduction}
The discovery of the accelerated expansion of the Universe has led to the development of a large number of theories of dark energy and modified gravity, a broad class of which are scalar-tensor theories involving a scalar degree of freedom. Many of the theories involving higher-order derivatives lead to Ostrogradski ghost instabilities \cite{Ostrogradsky:1850fid}, which can be avoided by introducing a set of degeneracy conditions \cite{Langlois:2015cwa, Motohashi:2016ftl, Motohashi:2017eya, Motohashi:2018pxg} which projects out the putative ghost degrees of freedom. Within this framework, the most  developed so far are the Degenerate Higher Order Scalar Tensor (DHOST) theories \cite{Langlois:2015cwa, Achour:2016rkg, BenAchour:2016fzp, Crisostomi:2016czh, Crisostomi:2018bsp, Bombacigno:2021bpk}, which  generalise the Horndeski \cite{Horndeski1974} and beyond-Horndeski theories \cite{Zumalacarregui:2013pma,Gleyzes:2014dya, Gleyzes:2014qga}. They are based on a single scalar field, and although their Euler-Lagrange equations are higher than second-order, they lead to a single propagating scalar degree of freedom, thus avoiding the Ostrogradski ghosts.  

The cosmological effect of the scalar field in these theories can be thought of as adding an effective energy-momentum tensor sourcing the Einstein-Hilbert action. It is well-known that stealth solutions where the scalar field becomes simply proportional to time  can be found in this context. The Einstein equations  can then lead  to  either Minkowski or de Sitter space-times, and the main effects of the scalar field  only appear at the perturbative level. 
However, it has also been shown that perturbations around these stealth solutions can be   strongly coupled \cite{Minamitsuji:2018vuw, deRham:2019gha, Motohashi:2019ymr, Khoury:2020aya} and can thus beyond the regime of validity of the EFT used to define the original model. In order to avoid this problem, the scordatura mechanism  can be at play \cite{Motohashi:2019ymr}, which  detunes the degeneracy condition in the EFT and  makes the perturbations weakly coupled, at the cost of adding a benign and apparent Ostrogradsky ghost whose mass is above the EFT cutoff scale. This scenario was first presented as ghost inflation \cite{ArkaniHamed:2003uy,ArkaniHamed:2003uz,Senatore:2004rj,Mukohyama:2005rw,Creminelli:2006xe}, i.e.  as a model of the early Universe. 

This class of theories have been used in the late Universe, but they can also be used in the early Universe to characterise inflation. The inflationary paradigm has been successfully developed starting from the late 1970s, and provides a source for the primordial quantum fluctuations \cite{Mukhanov:1981xt, Hawking:1982cz, Guth:1982ec, Starobinsky:1982ee} necessary to generate the matter perturbations leading to the appearance of structures in the Universe. The latest Cosmic Microwave background (CMB) experiments, such as \textit{Planck} \cite{Akrami:2018odb}, have been able to place the tightest constraints so far on the physics of inflation through the angular power spectrum of CMB anisotropies, and in particular on adiabatic scalar perturbations, showing a small departure from scale invariance. This can be quantified at the level of the power spectrum through 
\begin{equation}
\mathcal{P}_{\zeta} (k) =  \mathcal{P}_{\zeta} (k_*)  \left(\frac{k}{k_*}\right)^{n_s-1} \, ,
\end{equation}
where $k_*$ is a pivot scale and $n_s$ is the spectral index. One can consider also the variation of the spectral index with respect to the scale and thus express the power spectrum as a series,
\begin{equation}
\log \mathcal{P}_{\zeta} (k) =  \log \mathcal{P}_{\zeta} (k_*) 
+ \frac{1}{2} \frac{d \log n_s}{d \log k} \left(\frac{k}{k_*}\right)^2 
+ \frac{1}{6} \frac{d^2 \log n_s}{d \log k^2} \left(\frac{k}{k_*}\right)^3 + \ldots \, .
\end{equation}
These coefficients have been measured by \textit{Planck} with exquisite accuracy.

Scalar-tensor theories are able to provide a mechanism for producing such perturbations that are compatible with the latest measurements, and in this paper we show that this is the case for extensions of the derivative DHOST models where two types of perturbations are considered.
In the first case, we determine the power spectrum in the derivative scordatura DHOST theories, where the perturbation operators  are of the derivative-type  in the scalar field, showing that this will always lead to a scale-invariant power spectrum of curvature perturbations. As this excluded by the CMB data, we consider a second type of perturbations by a potential term, and we show that in this case there  will always be a departure from scale invariance. We  quantify this departure by determining the power spectrum, the spectral index and its first two derivatives for axion-like potentials, which break the shift symmetry in field space and could result from non-perturbative dynamics. Contrary to traditional inflationary models with polynomial interactions, here the mass and quartic interactions are small perturbations to the background and do not drive inflation. They are only responsible for the breaking of scale invariance in the power spectrum. Notice too that these polynomial interactions do not break the degeneracy condition and therefore do not reintroduce a ghost in the spectrum of the theory. Similarly the scale of inflation and the strong coupling scale of the theories are chosen such that the speed of sound squared is always positive, i.e. avoiding the strong coupling issue in the inflationary background. In section \ref{sec:scor}, we present the DHOST models of inflation at the background level and their perturbation by Scordurata terms. When both the DHOST and Scordurata operators are only derivative dependent, we find that the power spectrum of curvature perturbation is flat. In section \ref{sec:cos}, we apply the same method to calculate the power spectrum in the case of perturbation by potential terms. For axion-like potentials, we find that a power spectrum compatible with the \textit{Planck} 2018 data can be found. We also give orders of magnitude estimates for the non-Gaussianities and find that they could exceed the \textit{Planck} bounds if the tensor to scalar ratio is too small. As a result, these inflationary models are within reach of future experimental programmes, which would probe the tensor to scalar ratio down to values of order of $r\simeq 10^{-3}$ \cite{Hazumi:2019lys}. We conclude with two appendices where the scordurata coefficients leading to the Mukhanov-Sasaki equation are given in appendix  \ref{appendixLi} and the coefficients of the cubic operators in the curvature perturbation for axion-like models are given in appendix \ref{fnl:coef}.

\section{ DHOST theories and Scordatura perturbations}
\label{sec:scor}
\subsection{The models} 
In this section, we describe the formalism for determining the power spectrum of primordial curvature perturbations in the DHOST theories, using the methods developed in \cite{Gorji:2020bfl}. We restrict ourselves to the study of quadratic DHOST theories.

The most general action involving up to second-order interactions in the scalar field can be written as
\begin{eqnarray}\label{action-DHOST}
S = \int d^4 x \sqrt{-g} \Big[ F_0(\phi,X) + F_1(\phi,X) \Box \phi + F_2(\phi,X) R 
+ \sum_{i=1}^5 A_i(\phi,X) L_i \Big] \,,
\end{eqnarray}
where $X=g^{\nu\eta}\phi_{\nu}\phi_{\eta}$, with $\phi_{\nu}\equiv\nabla_{\nu}\phi$, and the sign convention is $(-,+,+,+)$.  $L_i$ are all the five possible Lagrangians quadratic in the field $\phi$ and $A_i(\phi,X)$ their corresponding amplitudes with
\begin{align}\label{DHOST-L2s}
L_1 &= \phi_{\nu\eta} \phi^{\nu\eta} , \hspace{1cm} L_2 = (\Box \phi)^2 ,
\hspace{1cm} L_3 = \Box\phi \, \phi_{\nu}\phi^{\nu\eta} \phi_{\eta} , 
\nonumber \\
L_4 &= \phi^{\nu} \phi_{\nu\eta} \phi^{\eta\lambda}\phi_{\lambda} , 
\hspace{1cm} L_5 = (\phi_{\nu}\phi^{\nu\eta}\phi_{\eta})^2 \,.
\end{align}
In the inflationary background, we will impose that $F_0>0$ and that it drives the expansion of the Universe. 
In order to be ghost-free and to satisfy the gravitational waves constraints, the functions $F_i$ and $A_i$ have to satisfy a set of degeneracy conditions \cite{Crisostomi:2017pjs, Crisostomi:2019yfo}.
In the following we simplify the setting and  assume that the functions $F_i$ and $A_i$ only depend on the kinetic term $X$. Imposing the degeneracy conditions, the DHOST action becomes
\begin{eqnarray}\label{action-dhost}
S_{\rm D} = \int d^4 x \sqrt{-g} \bigg[ F_0(X) + F_1(X) \Box\phi + F_2(X) R
+ \frac{6 F_{2,X}^2}{F_2} \phi^{\nu} \phi_{\nu\eta} \phi^{\eta\lambda}\phi_{\lambda} \bigg]\,.
\end{eqnarray}
{We note that the action presented above partially relies on gravitational wave decay into dark energy constraints \cite{Creminelli:2018xsv}. As we are dealing with an early universe scenario, this constraint might not be necessary. In what follows we choose a conservative path and maintain all the constraints derived on these models.}
We will consider small perturbations about the DHOST background, by adding a term, called the \textit{scordatura} correction, of the form of one of the $L_i$ corrections multiplied by a small parameter $\alpha$. We choose here the $L_2$ term, and we present the results for the other four in Appendix \ref{appendixLi}. We will also add simply a potential term as will be seen below in section \ref{sec:cos}. In the scordurata case, the total action is given by 
\begin{eqnarray}\label{action-scordatura}
S_{\rm g} = S_{\rm D} + S_{\rm S}  
\end{eqnarray}
where
\begin{equation}
\label{eq:scodL2}
S_{\rm S} = \int d^4 x \sqrt{-g} \bigg[ - \frac{\alpha}{2} \frac{(\Box\phi)^2}{M_S^2} \bigg] \,,    
\end{equation}
and we have assumed that the shift symmetry $\phi\to\phi +c$ is preserved for simplicity. Here $F_{2,X}=\partial F_2/\partial X$ and  $M_S$ is a mass scale related to the strong coupling scale of the EFT. $\alpha$ is a dimensionless parameter governing the strength of the scordatura term.  For $\alpha\neq0$ the scordatura term slightly breaks the degeneracy condition.
When perturbing by a potential interaction term, we will also assume that the coefficient is small enough to be treated as a small perturbation to the background cosmology driven by the DHOST action. 
Typically we will consider
\begin{equation}
    S_{\rm V} = -\int d^4 x \sqrt{-g}V(\phi)
\end{equation}
where $V(\phi)$ is an interaction, for instance of the form $V(\phi)= \mu^4 (\cos\frac{\phi}{f}-1)$ whose origin could be a non-perturbative breaking of the shift symmetry $\phi\to \phi+c$ like in the case of axions \cite{Marsh:2015xka}.

The equations of motion are determined by considering the variation of the action with respect to the metric and the scalar field. Before doing so, we first rewrite the action in terms of dimensionless coordinates and variables defined by:
\begin{equation}\label{coordinets-ch}
{\tilde t} \equiv \Lambda t \,, \hspace{1cm} {\tilde x}^i \equiv \Lambda x^i \,,
\end{equation}
\begin{equation}\label{dimensionless-couplings}
\phi \equiv M \, \varphi\,, \hspace{.5cm} X\equiv{M^2 \Lambda^2}{\mathrm x}\,, \hspace{.5cm} 
F_0 \equiv \Lambda^4 f_0 \,, \hspace{.5cm}
F_1 \equiv \frac{\Lambda^2}{M} f_1 \,, \hspace{.5cm} F_2 \equiv \Lambda^2 f_2 \,.
\end{equation}
where  we consider the models as low energy effective theories well below the Planck scale where quantum gravity effects should be considered. In essence, we assume that the DHOST theories could result from some ghost-free dynamics at high energy with the scordurata terms possibly appearing from quantum effects. The polynomial interactions on the other hand could result from the breaking of the shift symmetry by non-perturbative effects, in a way similar to what happens to axions.
In the following and at the background level of cosmology we will have 
$f_2={\cal O}(1)$ implying that $\Lambda\simeq m_{\rm Pl}$. In these units, time and space are measured in Planck length. The scale $M$ gives the typical excursion scale of the scalar field and to avoid large excursion in units of the Planck scale, we will require that $M\ll m_{\rm Pl}$. Moreover the expansion of the function $F_i$ in powers of ${\rm x}$  makes sense as long as $\vert X\vert \ll M^2\Lambda^2$ which defines the strong coupling scale
\begin{equation}
\mu_c= \sqrt{M\Lambda}
\end{equation}
which must also be $\mu_c\ll m_{\rm Pl}$.
In the following we shall simply treat the DHOST action and its perturbation as a low energy effective theory below $\mu_c$.

\subsection{DHOST Background}\label{sec-cos-BG}
In terms of the new dimensionless variables, the background FLRW (Friedmann-Lema\^itre-Robertson-Walker) metric takes the form  
\begin{equation}\label{metric-FRW-BG}
ds^2 = \Lambda^2 \Big[ - d{\tilde t}^2 + a({\tilde t})^2 \delta_{ij} d{\tilde x}^i d{\tilde x}^j \Big] \,,
\end{equation}
where $a({\tilde t})$ is the scale factor, ${\tilde t}$ and ${\tilde x}^i$, with $i=1,2,3$ are the dimensionless cosmic time and dimensionless spatial coordinates respectively. 
In terms of these new dimensionless quantities for the background geometry in the DHOST case, the two Einstein equations from the time and space components of the Einstein tensor $G_{00}$ and $G_{ii}$  are 
\begin{eqnarray}\label{Friedmann}
&&f_0 + 2 \dot{\varphi}^2 f_{0,{\mathrm x}} + 6 f_2 h^2
+ 12 \dot{\varphi}^2 f_{2,{\mathrm x}} (2 h^2 + \dot{h} )
- 12 \dot{\varphi} \ddot{\varphi} f_{2,{\mathrm x}} h
- 6  \dot{\varphi}^3 f_{1,{\mathrm x}} h
 \\ \nonumber
&& \qquad - 6 \dot{\varphi}^2 \frac{ f_{2,{\mathrm x}}^2}{f_2} 
\Big( \ddot{\varphi}^2 + 2 \dddot{\varphi} \dot{\varphi} + 6 h \dot{\varphi} \ddot{\varphi}
+ 2  (\dot{\varphi}\ddot{\varphi})^2 \Big( \frac{f_{2,{\mathrm x}}}{f_2} 
- 2 \frac{f_{2,{\mathrm x}{\mathrm x}}}{f_{2,{\mathrm x}}} \Big)
\Big)
= 0 \,,
\end{eqnarray}
\begin{eqnarray}\label{Raychuadhuri}
&&f_0 + 2f_2 (2\dot{h}+3h^2) - 4 f_{2,{\mathrm x}} \Big( (\dot{\varphi}\ddot{\varphi})\dot{} 
+ h \dot{\varphi} \ddot{\varphi} + 2 \dot{\varphi}^2 \ddot{\varphi}^2 
\Big(\frac{3f_{2,{\mathrm x}}}{4f_2} - \frac{f_{2,{\mathrm x}{\mathrm x}}}{f_{2,{\mathrm x}}} \Big) \Big) 
\nonumber \\ 
&& \qquad - 2  f_{1,{\mathrm x}} \dot{\varphi}^2 \ddot{\varphi} 
= 0 \,.
\end{eqnarray}
where a dot denotes a derivative with respect to the dimensionless cosmic time ${\tilde t}$, $h\equiv \dot{a}/a$ is the dimensionless Hubble parameter which is related to the standard Hubble parameter as $H = \Lambda h$.
Typically we expect inflation to be driven by an energy density below the scale $\mu_c$ where we expect that new physics will take place, as a result we shall impose
\begin{equation}
    h\ll 1 \, .
\end{equation}
It is convenient to 
consider the following change of variable
\begin{equation}\label{b}
 b \equiv \sqrt{f_2} \, a \,,
\end{equation}
and we define the following dimensionless quantities \cite{Bellini:2014fua, Gleyzes:2014rba, Langlois:2017mxy, Motohashi:2017gqb}
\begin{equation}\label{alpha-i}
\alpha_H \equiv - {\rm x} \frac{f_{2,{\rm x}}}{f_2}\,, \hspace{1cm}
\alpha_B \equiv \frac{1}{2} \frac{\dot{\varphi}\,{\rm x}}{h_b} \frac{f_{1,{\rm x}}}{f_2} + \alpha_H \,, \hspace{1cm}
\alpha_K \equiv - \frac{{\rm x}}{6h_b^2}\frac{f_{0,{\rm x}}}{f_2} + \alpha_H + \alpha_B \,,
\end{equation}
which are first order in derivative of the functions $f_i$.  The two Friedmann equations can then be expressed as 
\begin{eqnarray}\label{Friedmann-b}
&&f_0 + 6 f_2 \Big( h_b^2 (1 + 2\alpha_K) 
+ 2 \Big( \dot{h}_b + h_b \frac{\ddot{\varphi}}{\dot{\varphi}} \alpha_B \Big) \alpha_H \Big)
= 0 \,,
\end{eqnarray}
\begin{eqnarray}\label{Raychuadhuri-b}
f_0 + 2f_2 \Big( 3 h_b^2 + 2 \Big( \dot{h}_b + h_b \frac{\ddot{\varphi}}{\dot{\varphi}} \alpha_B \Big) \Big)
+ \frac{\alpha}{2}  \Big( (\Box\varphi)^2-2\dot{\varphi}(\Box\varphi)\dot{} \Big) 
= 0 \, ,
\end{eqnarray}
where 
\begin{equation}\label{Hubble-b}
h_b \equiv \frac{\dot{b}}{b} 
= h - \frac{\ddot{\varphi}}{\dot{\varphi}} \alpha_H 
\end{equation}
is the dimensionless Hubble parameter defined with respect to the new scale factor $b$.

\subsection{DHOST perturbations}\label{sec-Cos-pert}

We investigate cosmological perturbations about the background cosmology  in the comoving gauge where the scalar field perturbations are absent. The line element for the scalar perturbations is then given by
\begin{equation}\label{metric-FRW-Perturbations}
ds^2 = \Lambda^2 \Big( - ( 1 + 2 A ) d{\tilde t}^2 
+ 2 {\tilde \partial}_i B d{\tilde t} d{\tilde x}^i 
+ a^2 ( 1 + 2 \psi ) \delta_{ij} d{\tilde x}^i d{\tilde x}^j \Big) \,,
\end{equation}
where $(A,B,\psi)$ are scalar perturbations depending on the dimensionless coordinates $({\tilde t},{\tilde x}^i)$ and ${\tilde \partial}_i$ is the derivative with respect to ${\tilde x}^i$. 

Substituting (\ref{metric-FRW-Perturbations}) in (\ref{action-DHOST}), expanding the action up to the quadratic order in the perturbations and integrating by parts, the second order action for the DHOST theory is obtained 
\begin{equation}\label{action-SS-bare}
S^{(2)}_{\rm D} \equiv \int d{\tilde t} \slashed{d}^3{\tilde k} 
 {\tilde {\cal L}}_{\rm D}^{(2)}(\dot{\psi},\psi,\dot{A},A,B) \, ,
\end{equation}
where we define $\slashed{d}k= \frac{dk}{2\pi}$ and $\slashed{\delta}(k)= 2\pi \delta (k)$.
By considering the field redefinition
\begin{equation}\label{xi-def}
\zeta \equiv \psi + \alpha_H A \,,
\end{equation}
we can write the second order action as
\begin{equation}\label{action-SS-bare-xi}
S^{(2)}_{\rm D} = \int d{\tilde t} \slashed{d}^3{\tilde k} M^4 
 {\tilde {\cal L}}_{\rm D}^{(2)}(\dot{\zeta},\zeta,A,B) \, ,
\end{equation}
where\footnote{In Fourier space, quadratic terms should be understood as $\zeta^2\equiv \zeta(-\vec k) \zeta(\vec k)$}
\begin{eqnarray}\label{Lagrangian-DHOST}
{\tilde {\cal L}}_{\rm D}^{(2)} &=& 2 f_2 \Big(
- 3 a^3 \dot{\zeta}^2 + 6 a^3 h_b (1+ \alpha_B) \dot{\zeta} A
- 2 a {\tilde k}^2 \dot{\zeta} B
+ a {\tilde k}^2 \zeta^2 \\ \nonumber
&& \hspace{1cm} + 2 a (1+ \alpha_H) {\tilde k}^2\zeta A 
- 3 a^3 h_b^2 \beta_K A^2 + 2 a h_b (1+\alpha_B) {\tilde k}^2 AB \Big) \,,
\end{eqnarray}
and we have defined the second order dimensionless parameters 
\begin{eqnarray}\label{beta-K}
&&\beta_K \equiv - \frac{{\rm x}^2}{3} \frac{f_{0,{\rm x}{\rm x}} }{h_b^2 f_2} 
+ (1-\alpha_H) (1+3 \alpha_B) + \beta_B 
+ \frac{(1 + 6 \alpha_H - 3 \alpha_H^2) \alpha_K
- 2 ( 2 - 6 \alpha_H + 3 \alpha_K ) \beta_H}{1-3 \alpha_H} \,,
\nonumber \\
&& \beta_B \equiv  \dot{\varphi}\,{\rm x}^2 \frac{f_{1,{\rm x}{\rm x}}}{h_b f_2} \,,
\hspace{1cm}
\beta_H \equiv {\rm x}^2 \frac{f_{2,{\rm x}{\rm x}}}{ f_2}
\,.
\end{eqnarray}
The variable $\zeta$ is the \textit{comoving curvature perturbation} \cite{Riotto2002}. 
Notice that the two fields
$A$ and $B$ can be treated as Lagrange multipliers and by varying the Lagrangian (\ref{Lagrangian-DHOST}) with respect to $A$ and $B$ and then solving the resulting equations we get
\begin{equation}\label{Phi-B-sol}
A = \frac{1}{1+\alpha_B} \frac{\dot{\zeta}}{h_b} \,, \hspace{1cm}
B = - 3 \bigg[ 1 - \frac{\beta_K}{(1+\alpha_B)^2} \bigg] \frac{a^2}{{\tilde k}^2} \, \dot{\zeta}
- \frac{1+\alpha_H}{1+\alpha_B} \, \frac{\zeta}{h_b} \,.
\end{equation}
Substituting the above results into the DHOST Lagrangian, we obtain \cite{Crisostomi:2018bsp}
\begin{equation}\label{Lagrangian-DHOST-red}
{\tilde {\cal L}}_{\rm D}^{(2)}
= a^3 f_2 \bigg( \bar{\cal A} \, \dot{\zeta}^2 - \bar{\cal B} \, \frac{{\tilde k}^2}{a^2} \zeta^2
\bigg) \,,
\end{equation}
where
\begin{eqnarray}\label{coefficients-AD-BD}
\bar{\cal A} = 6 \bigg[ 1 - \frac{\beta_K}{(1+\alpha_B)^2} \bigg] \,, \hspace{1cm}
 \bar{\cal B} = - 2 \bigg[ 1 - \frac{1}{a f_2} \frac{d}{d{\tilde t}} 
\bigg(\frac{af_2}{h_b}\frac{1+\alpha_H}{1+\alpha_B}\bigg) \bigg] \,.
\end{eqnarray}
The Euler-Lagrange equations for $\zeta$ from (\ref{Lagrangian-DHOST-red}), give the equation of motion for $\zeta$,
\begin{equation}\label{EoM-zeta}
\ddot{\zeta} + \Big( 3 h + \frac{d}{d{\tilde t}} \ln(f_2 \bar{\cal A} ) \Big) \dot{\zeta} 
+ \Big( \frac{\bar{c}_{\rm s} {\tilde k}}{a}\Big)^2 \zeta = 0 \,.
\end{equation}
where $\bar{c}_{\rm s}^2=\frac{{\cal B} }{{\cal A} }$.
We will discuss the sign of the speed of sound squared below when we specify the inflationary background.

\subsection{Scordatura corrections}
\label{sec:scod_pert}

The scordatura action can be similarly expanded to second order, and expressed in terms of the perturbation variables as:
\begin{equation}\label{action-SS-bare-xi-all}
S^{(2)}_{\rm S} = \int d{\tilde t} \slashed{d}^3{\tilde k}  \alpha
  {\tilde {\cal L}}_{\rm S}^{(2)}(\dot{\zeta},\zeta,\dot{A},A,B )  \,,
\end{equation}
where we have redefined $ \alpha \frac{M^2}{M_S^2}\to \alpha$
The Lagrangian then takes the following form:
\begin{align}\label{Lagrangian-scordatura}
{\tilde {\cal L}}_{\rm S}^{(2)} &= \frac{1}{2} \Big(
{\bar k}_{11} \dot{\zeta}^2 + {\bar k}_{22} \dot{A}^2 + 2 {\bar k}_{12} \dot{\zeta} \dot{A}
+ 2 \dot{\zeta} ({\bar n}_{12} A + {\bar n}_{13} {\tilde k}^2 B) + 2 {\bar n}_{23} {\tilde k}^2 \dot{A} B - {\bar m}_{11} \zeta^2
\nonumber \\
&   - 2 {\bar m}_{12} \zeta A - {\bar m}_{22} A^2 - {\bar m}_{22{\rm s}}{\tilde k}^2 A^2 
- 2 {\bar m}_{23} {\tilde k}^2 A B + {\bar m}_{33} {\tilde k}^2 B^2 + {\bar m}_{33{\rm s}} {\tilde k}^4 B^2
 \Big) \,,
\end{align}
where the various coefficients are given by
\begin{align}\label{matrices-components-scordatura}
& {\bar k}_{11} = 9 a^3 {\rm x} , \hspace{2.1cm}
{\bar k}_{12} = - 3 a^3 {\rm x} (1+3\alpha_H) , \hspace{2.1cm} 
{\bar k}_{22} = a^3 {\rm x} (1+3\alpha_H)^2 , \\ \nonumber
& {\bar n}_{12} = - 6 a^3 \big( 3{\rm x} h_b 
- (1+6\alpha_H + 3 \alpha_{H}^2-3\beta_H) \dot{\varphi} \ddot{\varphi} \big) , \hspace{.5cm}
{\bar n}_{13} = 3 a {\rm x} , \hspace{.5cm} 
{\bar n}_{23} = - a {\rm x} (1+3\alpha_H) , \\[10pt] \nonumber
& {\bar m}_{11}  = - \frac{3}{2} a^3 \Big( - 3 {\rm x} ( 2\dot{h}_b  + 3 h_b^2 ) 
+ 6 ( 2 + 3 \alpha_H ) h_b \dot{\varphi} \ddot{\varphi}
\\ \nonumber 
& \hspace{2.2cm} + ( 1 + 18 \alpha_H + 21 \alpha_H^2 
- 12 \beta_H ) \ddot{\varphi}^2 
+ 2 ( 1 + 3 \alpha_H ) \dot{\varphi} \dddot{\varphi} \Big) , 
\\ \nonumber
& {\bar m}_{12} = - \frac{3}{2} a^3 \Big( 
- 9 ( 1 - \alpha_H ) h_b^2 {\rm x} + 6 ( 1 + \alpha_H ) {\rm x} \dot{h}_b
 + 6 (1-3\alpha_H) \alpha_H h_b \dot{\varphi} \ddot{\varphi}
\\ \nonumber
& \hspace{2.25cm} + \big( 1 - \big( 7 + 21 \alpha_H + 21 \alpha_H^2 
- 12 \beta_H \big) \alpha_H + 12 \beta_H \big) \ddot{\varphi}^2
- 2 ( 1 + \alpha_H ) ( 1 + 3 \alpha_H ) \dot{\varphi}  \dddot{\varphi}
\Big) , \\ \nonumber
& {\bar m}_{22} = 
- \frac{1}{2} a^3 \Big( 9 \big( 5 - 6 \alpha_H - 3 \alpha_H^2 \big) {\rm x} h_b^2 
- 6 \big( 5 + 12 \alpha_H + 3 \alpha_H^2 \big) {\rm x} \dot{h}_b 
\\ \nonumber
& 
\hspace{2.3cm} - 18 h_b \big( ( 1 - 12 \alpha_H - 9 \alpha_H^2
+ 6 \beta_H ) \alpha_H - 2 \beta_H \big) \dot{\varphi} \ddot{\varphi} 
\\ \nonumber
& \hspace{2.3cm} + ( 3 \gamma_H - 5) \ddot{\varphi}^2  
+ 2 ( 1 + 3 \alpha_H ) 
\big( 5 + 18 \alpha_H + 9 \alpha_H^2 
- 6 \beta_H \big) \dot{\varphi} \dddot{\varphi} \Big) ,
\\ \nonumber
& {\bar m}_{22{\rm s}} =0 \\ \nonumber
& {\bar m}_{23} = - 2 a \Big( - 3 h_b {\rm x} 
+ \big( 1 + 6 \alpha_H + 3 \alpha_H^2 - 3 \beta_H \big) \dot{\varphi} \ddot{\varphi}  \Big) \,, \hspace{1cm}
{\bar m}_{33{\rm s}} = \frac{{\rm x}}{a} \,, \nonumber \\ \nonumber
& {\bar m}_{33} = \frac{1}{2} a \Big( - 9 h_b^2 {\rm x} + 6 {\rm x} \dot{h}_b 
+ 18 h_b \alpha_H \dot{\varphi} \ddot{\varphi} 
+ \big( 1 - 6 \alpha_H -3 \alpha_H^2 
+ 12 \beta_H \big) \ddot{\varphi}^2 - 2 ( 1 + 3 \alpha_H ) \dot{\varphi} \dddot{\varphi} \Big) \,,
\end{align}
Notice  that we have corrected the expressions and added a new  term $({\bar m}_{22{\rm s}}{\tilde k}^2 A^2)$ compared to the results found in Ref. \cite{Gorji:2020bfl}\footnote{ We thank Mohammad Ali Gorji for correspondence and agreeing with our corrections.}. Although this does not change the result for the $L_2$ scordatura term, this  will appear in the $L_4$ scordatura term (see Appendix \ref{appendixLi}). 

From (\ref{Lagrangian-scordatura}) we see that it is not possible to remove the time derivative of the field $A$ using integration by parts, meaning that there will always be  an Ostrogradsky ghost in this theory. In order to proceed from this point, we see that substituting the DHOST solution for $B$ yields an infrared divergence as ${\tilde k} \to 0$. As  there is no physical divergence in this limit  we proceed as follows \cite{Gorji:2020bfl}:
\begin{itemize}
    \item for $A$, we use the DHOST solution;
    \item for $\dot{A}$, we differentiate the DHOST solution (\ref{Phi-B-sol}) and using (\ref{EoM-zeta}), we find
    \begin{eqnarray}\label{Phidot}
\dot{A} = - \frac{1}{h_b(1+\alpha_B)} \bigg(
\frac{d}{d{\tilde t}} \Big[ \ln \big(h_b (1+\alpha_B) a^3 f_2 \bar{\cal A} \big)\Big] \, \dot{\zeta}
+ \Big( \frac{\bar{c}_{\rm s} {\tilde k}}{a}\Big)^2 \zeta 
\bigg) \, ;
\end{eqnarray}

\item for $B$, we notice that only the ${\bar m}_{33}$ term contributes to the divergence and hence we solve the Euler-Lagrange equations for $B$ by adding the term (${\bar m}_{33} {\tilde k}^2 B^2/2$) to the background Lagrangian, obtaining
\begin{equation}\label{B-sol}
B = - 3 \bigg[ 1 - \frac{\beta_K}{(1+\alpha_B)^2} \bigg] \frac{a^2\dot{\zeta}}{
{\tilde k}^2+\alpha k_{\rm IR}^2}
- \frac{1+\alpha_H}{1+\alpha_B} \, \frac{\zeta}{h_b} \,; \hspace{1cm} 
k_{\rm IR}^2 \equiv \frac{3 a}{4f_2} \frac{\beta_K{\bar m}_{33}}{(1+\alpha_B)^2} \,.
\end{equation}
    
\end{itemize}
Notice that this provides us with an infrared cut-off scale $k_{\rm IR}$ which depends on the perturbation coupling $\alpha$, acting as a resummation of infrared sensitive effects in the propagator $1/k^2$. 
Hence, replacing these three quantities, and integrating by parts, we get
\begin{equation}\label{Lagrangian-scordatura-red}
{\tilde {\cal L}}_{\rm S}^{(2)} = \frac{a^3}{2} \bigg[ \Big( {\cal A}_1 
+ \frac{a^2 {\cal A}_{2}}{{\tilde k}^2+\alpha {k}_{\rm IR}^2} \Big) \, \dot{\zeta}^2 
- \Big( {\cal B}_{1} \Big(\frac{{\tilde k}}{a}\Big)^2 + {\cal B}_{2}
\Big(\frac{{\tilde k}}{a}\Big)^4 + {\cal M}\Big) \,\zeta^2
\bigg] \,,
\end{equation}
where we have defined five new coefficients ${\cal A}_{1}, {\cal A}_{2}, {\cal B}_{1}, {\cal B}_{2}$, and ${\cal M}$ which will be explicitly spelt out below. 

The total quadratic  Lagrangian for scalar perturbations can  thus be expressed as
\begin{equation}\label{Lagrangian-red}
{\tilde {\cal L}}^{(2)}_{\rm g}
= a^3 f_2 \, {\cal K} \bigg[ \dot{\zeta}^2 
- \Big( c_{\rm s}^2 ({\tilde k})\frac{{\tilde k}^2}{a^2} + \alpha  m^2 \Big) \zeta^2
\bigg] \,,
\end{equation}
where we have defined the effective scale-dependent sound speed square  as
\begin{eqnarray}\label{cs2}
c_{\rm s}^2 ({\tilde k}) \equiv \bar{c}_{\rm s}^2 
+ \frac{\alpha}{2f_2} \bigg(
\frac{{\cal B}_{1}}{\bar{\cal A}} - \bar{c}_{\rm s}^2 \frac{{\cal A}_{1}}{{\cal A}}
+ \Big(\frac{{\tilde k}}{a}\Big)^{2} \frac{{\cal B}_{2}}{\bar{{\cal A}}}
\bigg) \,,
\end{eqnarray}
the mass term
\begin{eqnarray}\label{mass}
m^2 \equiv \frac{1}{2f_2}
\bigg(
\frac{{\cal M}}{\bar{\cal A}} 
- \bar{c}_{\rm s}^2 \frac{{\cal A}_{2}}{\bar{\cal A}}
\bigg) \,,
\end{eqnarray}
and the kinetic term coefficient
\begin{equation}\label{ghost}
{\cal K} \equiv \bar{\cal A} 
\bigg( 1 + \frac{\alpha}{2f_2} \Big( \frac{{\cal A}_{1}}{\bar{\cal A}} 
+ \frac{a^2}{{\tilde k}^2+\alpha k_{\rm IR}^2} \frac{{\cal A}_{2}}{\bar{\cal A}} \Big)
\bigg) \,.
\end{equation}
Notice that the mass appears due to the scodurata perturbation. The rescaling by ${\cal K}$ will be at the origin of the rescaling to obtain the Mukhanov-Sasaki variable $v$. 

\subsection{de Sitter background cosmology}\label{sec-stealth-DE}

The DHOST equations of motion have inflationary solutions obtained by choosing time as the scalar field, i.e. 
\begin{equation}\label{stealth-sol}
\varphi({\tilde t}) = {\tilde t}\,, \hspace{1cm} {\mathrm x} = -1 \,,
\end{equation}
which after substituting in (\ref{Friedmann}) and (\ref{Raychuadhuri}) and some manipulations is valid provided the Lagrangian satisfies the identities
\begin{eqnarray}
&& f_0 +  6 h_{\rm dS}^2  f_2  = 0 \,, \label{Stealth-Raychuadhuri}
\\
&& f_{0,{\mathrm x}} + 3 h_{\rm dS} ( 4 h_{\rm dS} f_{2,{\mathrm x}} -  f_{1,{\mathrm x}} ) = 0 \,. \label{Stealth-Friedmann}
\end{eqnarray}
The coefficients are evaluated for $\mathrm x=-1$ and are therefore constants.
Here, $h_{\rm dS}$ is the constant Hubble parameter that arises for the stealth solution (\ref{stealth-sol}), in particular  this implies that  $\alpha_K=0$ and 
\begin{equation}\label{Stealth-Hubble}
h_{\rm dS} = \sqrt{\frac{-f_0}{6 f_2}} \,. 
\end{equation}
Hence, we must impose that $f_0<0$ corresponding to a positive energy density and $f_2>0$, i.e. the sign of the Einstein-Hilbert term is positive guaranteeing the absence of ghosts in the gravitational wave spectrum of the theory. The coefficients determining the Lagrangian for the curvature perturbations are given by 
\begin{align}\label{coefficients-AD-BD-ST}
&\bar{\cal B} = - 2 \frac{ \alpha_B - \alpha_H }{1+ \alpha_B} \,, \\
&\bar{\cal A} = 2 \frac{ \left( 4 f_2 f_{0,{\rm x}{\rm x}}
+3  f_{1,{\rm x}}^2 + 6  h_{\rm dS} \big( f_2 \big( f_{1,{\rm x}} - 2 f_{1,{\rm x}{\rm x}} \big) 
- 5 f_{1,{\rm x}} f_{2{\rm x}} \big) + 48 h_{\rm dS}^2 \left( f_2 f_{2,{\rm x}{\rm x}}
+ \left( f_{2,{\rm x}} \right)^2 \right) \right)}{ \big(  f_{1,{\rm x}}
-2 h_{\rm dS} \left( f_2 + f_{2,{\rm x}} \right)\big)^2} \,,\nonumber \\
\end{align}
and the speed of sound squared is given by 
\begin{align}\label{Stealth-cs2}
\bar{c}_{\rm s}^2 &= \frac{ f_{1,{\rm x}} \left(2 h_{\rm dS} \left(f_{2,{\rm x}} + f_2 \right)- f_{1,{\rm x}} \right)}{4 f_2 f_{0,{\rm x}{\rm x}}+3  f_{1,{\rm x}}^2+6 h_{\rm dS} 
	 \left(f_{1,{\rm x}} \left(f_2-5 f_{2,{\rm x}} \right)-2 f_2 f_{1,{\rm x}{\rm x}}\right) 
	+ 48 h_{\rm dS} ^2 \left( f_2 f_{2,{\rm x}{\rm x}} + f_{2,{\rm x}}^2 \right)} \nonumber\\ 
 & = -\frac{(1+\alpha_B)(\alpha_B-\alpha_H)}{3 (1+2\alpha_B+\alpha_B^2-\beta_K)}
	\, .
\end{align}
As can be seen in the first expression, the speed of sound squared can be negative in particular when the limit $h_{\rm dS}\to 0$ is taken.   In this case the term in $-f_{1,{\rm x}}^2$ implies that $\bar{c}_{\rm s}^2<0$ \cite{Motohashi:2019ymr}. In the following, we shall concentrate on a different limit where $h_{\rm dS}\ne 0$ and $h_{\rm dS}\lesssim 1$ to make sure that inflation is not driven by an energy density beyond the Planck scale. We will find that this allows one to get $\bar{c}_{\rm s}^2>0$, i.e. 
the speed of sound squared becomes positive only in a de Sitter background.
In practice, we use the parameterisation in terms of the $\alpha$'s and $\beta$'s coefficients as they are more appropriate to parameterise the curvature perturbations. 
For the scordatura corrections, we obtain the following coefficients:
\begin{align}\label{coefficients-AS-BS-CS}
{\cal A}_{1} &= - \frac{9}{2 (1+ \alpha_B)^4} ( \alpha_B^2 \left(15 \alpha_H^2+6 \alpha_H+7\right)+2 \alpha_B \left(15 \alpha_H^2+6 \alpha_H (\beta_K+1)-2 \beta_K+7\right) \nonumber\\
& \qquad + 15 \alpha_H^2+6 \alpha_H (2 \beta_K+1)+2 \beta_K^2-4 \beta_K+7
) \nonumber \\[5pt]
{\cal A}_{2} &= \frac{81}{2} h_{\rm dS}^2 
\bigg[ 1 - \frac{\beta_K}{(1+\alpha_B)^2} \bigg]^2 \,, \hspace{1cm} 
k_{\rm IR}^2 = \frac{27}{8} \frac{a^2\beta_K}{f_2(1+\alpha_B)^2} 
\nonumber \\[10pt]
{\cal B}_{1} &= \frac{1}{2 (\alpha_B+1)^3 \left(\alpha_B^2+2 \alpha_B-\beta_K+1\right)} \left[\alpha_B^2 \left(-18 \alpha_H^3+51 \alpha_H^2+4 \alpha_H (3 \beta_K+1)+8 \beta_K-41\right) \right. \nonumber \\[5pt]
& \quad  +\alpha_B^3 \left(27 \alpha_H^2+6 \alpha_H-13\right)+\alpha_B \left(-3 \alpha_H^2 (5 \beta_K-7)-36 \alpha_H^3+2 \alpha_H (11 \beta_K-5)+29 \beta_K-43\right) \nonumber \\
& \quad \left. -3 \alpha_H^2 (5 \beta_K+1)-18 \alpha_H^3-2 \alpha_H \left(3 \beta_K^2-5 \beta_K+4\right)-3 \left(2 \beta_K^2-7 \beta_K+5\right) \right] \nonumber \\ 
{\cal B}_{2} &= \frac{1}{9 h_{\rm dS}^2} \left[ \frac{\alpha_B^2 (6 \alpha_H+4)+\alpha_B \left(-3 \alpha_H^2+8 \alpha_H+7\right)-3 \alpha_H^2+\alpha_H (2-3 \beta_K)-3 \beta_K+3}{(\alpha_B+1)  \left(\alpha_B^2+2 \alpha_B-\beta_K+1\right)} \right]^2 \nonumber \\[10pt]
{\cal M} &= -\frac{27}{2} h_{\rm dS}^2 \frac{ ( \alpha_B - 5) ( 1 + \alpha_B )^2+3 ( 1 + \alpha_H ) \beta_K }{( 1 + \alpha_B )^3} \,.
\end{align}
As we shall see, their exact expressions are not crucial as the curvature perturbations in these models will be shown to be characterised by a scale invariant spectrum. 

\subsection{Quantisation}

\begin{figure}[!h]
\centering
\includegraphics[width=0.49\linewidth]{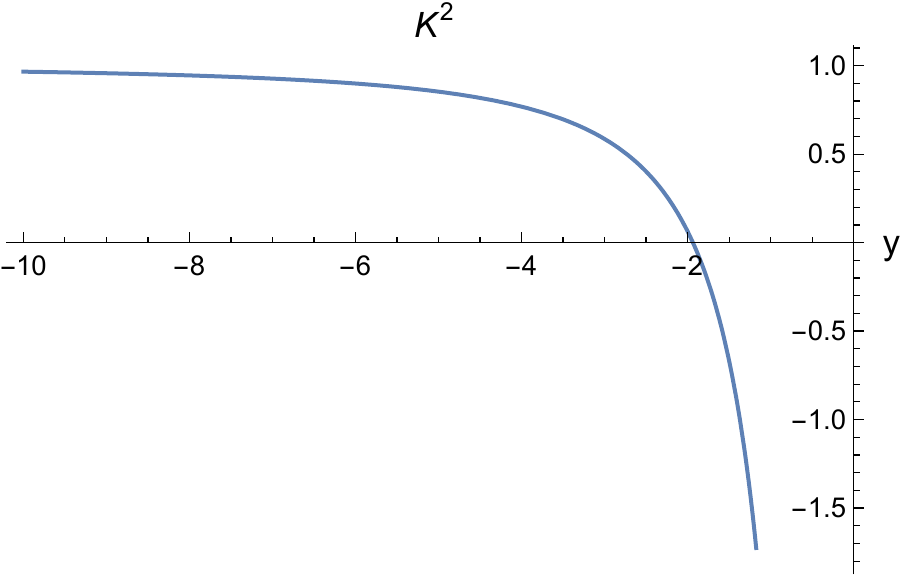} 
\includegraphics[width=0.49\linewidth]{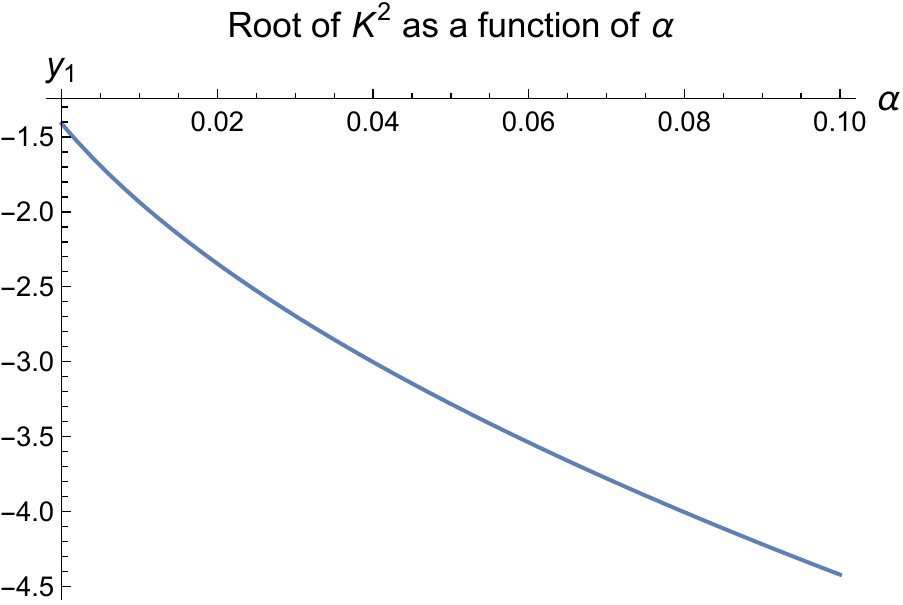} 
\caption{The function $K^2(y)$ for $\alpha=10^{-2}$ and $\alpha_H = 1.04$, $\alpha_B = 1$, $\beta_K = 3.97343$, $h_{\mathrm{ds}} =  10^{-2}$, $c = 1$, $ = 5.12 \times 10^{-13}$, $f_2= 1$ (left) showing that it only has one negative root. The position of the negative root varies with $\alpha$ as can be seen in the right panel (using the same parameters).}
\label{fig:K2}
\end{figure}

Now that we have reduced the action for the curvature perturbation to a canonical form, we can consider the creation of primordial fluctuations from the Bunch-Davies vacuum in these models. It is convenient to parameterise
the de Sitter phase in the early Universe by the scale factor in conformal time
\begin{equation}
\label{eq:a}
a(\eta)=-\frac{1}{h_{\mathrm{dS}} \eta}   \,
\end{equation}
where $\eta$ is the conformal time whose range is   $-\infty < \eta < 0$. We can re-express the action Eq. (\ref{Lagrangian-red}) in conformal time as
\begin{equation}\label{Lagrangian-red-conf}
S^{(2)}_{\rm g}
= \int d\eta \slashed{d}^3 \tilde k  z^2  \,  \bigg[ \zeta'^2 
- a^2\Big( c_{\rm s}^2 ({\tilde k})\frac{{\tilde k}^2}{a^2} + \alpha  m^2 \Big) \zeta^2
\bigg] \,,
\end{equation}
where $z^2=a^2 f_2 \cal{K}$ and hence 
\begin{equation}
z=\frac{1}{h_{\mathrm{dS}} \eta} \sqrt{f_2 {\cal A}} \sqrt{\bigg( 1 + \frac{\alpha }{2f_2} \Big( \frac{{\cal A}_{1}}{\bar{\cal A}} 
+ \frac{1}{h_{\mathrm{dS}}^2 \eta^2}\frac{1}{{\tilde k}^2+\alpha k_{\rm IR}^2} \frac{{\cal A}_{2}}{\bar{\cal A}} \Big)
\bigg)} \, .
\end{equation}
We proceed in the standard way \cite{Peter:2013avv} to quantise the scalar field and hence to calculate the power spectrum of $\zeta$. We first define the modified Mukhanov-Sasaki variable as $v = z \zeta$, which satisfies the following equation of motion 
\begin{equation}
    v''+\Big[ a^2\Big( c_{\rm s}^2 ({\tilde k})\frac{{\tilde k}^2}{a^2} + \alpha   m^2 \Big)-\frac{z''}{z} \Big] v=0 \, .
\label{eq:ms}    
\end{equation}
Then, calculating $z''/z$, keeping the term ${\tilde k}^2+\alpha k_{\rm IR}^2$ corresponding to the removal of the infrared divergence, but otherwise expanding up to first order in $\alpha$, we obtain
\begin{equation}
\frac{z''}{z}=\frac{2}{\eta^2}+\frac{5 {\cal A}_2 }{2 f_2 {\cal A} h_{\mathrm{dS}}^2  \eta^4 ({\tilde k}^2+\alpha k_{\rm IR}^2)}  \alpha \,. 
\end{equation}
Then, to  first order in $\alpha$, the coefficient of $v$ in Eq. (\ref{eq:ms}) can be written  as
\begin{align}
&\left[\bar{c}_{\rm s}^2 {\tilde k}^2-\frac{2}{\eta^2}\right]
+ \alpha \left[- \frac{5 {\cal A}_2 }{2 f_2 \bar{{\cal A}} h_{\mathrm{dS}}^2  \eta^4 ({\tilde k}^2+\alpha k_{\rm IR}^2)} + \frac{m^2}{h_{\mathrm{dS}}^2 \eta^2} +\frac{ k^2}{2 f_2} \left( \frac{{\cal B}_1}{\bar{{\cal A}}}- \bar{c}_{\rm s}^2 \frac{{\cal A}_1}{\bar{{\cal A}}}  \right) + \frac{1}{2 f_2}k^4 h_{\mathrm{dS}}^2 \eta^2 \frac{{\cal B}_2}{\bar{{\cal A}}}\right] \,. 
\end{align}
Now after  performing the  change of variable $y={\tilde k} \eta$, eq. (\ref{eq:ms}) can be expressed as 
\begin{equation}
    v''(y)+K^2(y) v(y)=0 \, ,
\label{eq:vxx}    
\end{equation}
where we have defined the momentum variable
\begin{align}
K^2(y)=\bar{c}_{\rm s}^2 &-\frac{2}{y^2}
+ \alpha \left[- \frac{5 {\cal A}_2 }{2 f_2 \bar{{\cal A}} h_{\mathrm{dS}}^2   (y^4+\alpha y^2 \frac{27 \beta_K }{8 f_2 (1+\alpha_B)^2})} + \frac{ m^2}{h_{\mathrm{dS}}^2 y^2} \right.\nonumber \\
&+\left.\frac{1}{2 f_2} \left( \frac{{\cal B}_1}{\bar{{\cal A}}}- \bar{c}_{\rm s}^2 \frac{{\cal A}_1}{\bar{{\cal A}}}  \right) + \frac{1}{2 f_2} h_{\mathrm{dS}}^2 y^2 \frac{{\cal B}_2}{\bar{{\cal A}}}\right]  \,.  
\label{K2}
\end{align}
In the following, we will focus on cases where the momentum function $K(y)$ admits a zero on the negative real axis. This is what happens in de Sitter case for a massless scalar field where $K(y_1)=0$ for $y_1= - \sqrt 2$.
Using the notation,
\begin{align}
b_1 &= \frac{5 {\cal A}_2 }{2 f_2 \bar{{\cal A}} h_{\mathrm{dS}}^2} \,,\nonumber \\
c_1 &= \frac{27 \beta_K }{8 f_2 (1+\alpha_B)^2} \,,\nonumber \\
d_1 &= \frac{ m^2}{h_{\mathrm{dS}}^2 } \,,\nonumber \\
e_1 &= \frac{1}{2 f_2} \left( \frac{{\cal B}_1}{\bar{{\cal A}}}- \bar{c}_{\rm s}^2 \frac{{\cal A}_1}{\bar{{\cal A}}}  \right) \,, \nonumber \\
f_1 &= \frac{1}{2 f_2} h_{\mathrm{dS}}^2  \frac{{\cal B}_2}{\bar{{\cal A}}} \,,
\end{align}
we can simplify 
\begin{equation}
K^2(y)=\bar{c}_{\rm s}^2 -\frac{2}{y^2}
+ \alpha \left[- \frac{b_1}{ (y^4+\alpha c_1 y^2)} + \frac{d_1}{y^2} +e_1 + f_1 y^2 \right] \,.
\end{equation}
As $d_1>0, f_1>0$, and assuming $b_1,c_1>0$, we find that 
\begin{align}
&\lim_{y\to-\infty} K^2(y) = + \infty \,,\\
&\lim_{y\to 0} K^2(y) = - \infty
\end{align}
for $\alpha \ll 1$, hence 
$K^2(y)=0$ has a unique root in the interval $(-\infty, 0)$, as  can be seen in the left panel of Fig. \ref{fig:K2} (for $\alpha=10^{-2})$. In this figure we have chosen the DHOST parameters as $\alpha_H = 1.04$, $\alpha_B = 1$, $\beta_K = 3.97343$, $h_{\mathrm{ds}} =  10^{-2}$, $c = 1$ and  $f_2= 1$. 

In the absence of the scordatura perturbations, we can determine the exact solution for the scalar power spectrum at the horizon $y_H=-1$ \cite{Garriga:1999vw} using 
the exact solution for the Mukhanov-Sasaki variable 
\begin{align}
\label{DHOST:scalar}
v({\tilde k},y)&=\frac{1}{\sqrt{2 \bar{c}_s {\tilde k}}}\left(1-\frac{i}{\bar{c}_s y} \right) \exp(-i \bar{c}_s y) \,,\\
z^2({\tilde k},y)&=\frac{6 {\tilde k}^2 f_2}{h_{\mathrm{dS}}^2 y^2} \left(1-\frac{\beta_K}{(1+\alpha_B)^2} \right) \,,
\end{align}
which converges to the appropriate Bunch-Davies vacuum when $\eta \to -\infty$. Notice that the vacuum state depends on $\bar c_s >0$. 
As can be seen the power spectrum at $y$ is given by
\begin{align}
\mathcal{P}_{\zeta}({\tilde k},y)&=\frac{{\tilde k}^3}{2 \pi^2} \left|\frac{v}{z} \right|^2 = \frac{h_{\mathrm{dS}}^2 y^2 \left(1+\frac{1}{\bar{c}_s^2 y^2} \right)}{12 \pi^2 f_2 \bar{c}_s \left(1-\frac{\beta_K}{(1+\alpha_B)^2} \right) } \, ,
\end{align}
which becomes scale-invariant, i.e. independent of $k$ when evaluated at horizon crossing for $y_H=-1$,
where $\bar{c}_s$ is given in Eq. (\ref{Stealth-cs2}).
 Notice that the power spectrum mainly depends on the values of $h_{\rm dS}$,\ $f_2$ and $\bar c_s$. In particular we will see below that the COBE normalisation of the power spectrum will fix this combination of parameters, in particular the value of $h_{\rm dS}$ will have to be much lower than unity.

As no exact solution can be found when $\alpha\ne 0$  we apply the improved WKB method using matched asymptotics as described in Chapter 5 of Ref. \cite{weinberg_2015}. This will allow us to find the solution of Eq. (\ref{eq:vxx}) in the vicinity of the unique negative root of $K(y)$, $y_1\sim y_H$, i.e. close to horizon crossing. In the right panel of Fig. \ref{fig:K2} we show how the root of the function $K(y)$ varies with $\alpha$, starting at $-\sqrt{2}$ when $\alpha=0$.
It will prove useful to define the variable
\begin{equation}
\tilde{\phi}(y) = \int_{y}^{y_1} K(y') dy'.
\end{equation}
Around $y_1$, where $K^2(y)>0$ to the left of the turning point,  we can expand the function $K(y)$ to linear order implying that 
\begin{equation}
K(y)=\beta_E \sqrt{y_1-y}    
\end{equation}
where 
\begin{equation}
\beta_E=\sqrt{-(K^2)'(y_1)}.   
\end{equation}
This expansion is valid in the interval $y_1-\delta_E \lesssim y \le y_1$, where
\begin{equation}
\delta_E = \left|\frac{2 (K^2)'(y_1)}{(K^2)''(y_1)}   \right|
\end{equation}
corresponding to neglecting the quadratic terms in $K^2$ around the turning point. 
In this interval we have 
\begin{equation}
\tilde{\phi}(y) \simeq  \frac{2 \beta_E}{3} (y_1-y)^{3/2}\, .
\end{equation}
Then Eq. (\ref{eq:vxx}) can be expressed around $y_1$ as
\begin{equation}
    \frac{d^2 v}{d \tilde{\phi}^2} + \frac{1}{3 \tilde{\phi}} \frac{dv}{d \tilde{\phi}} + v = 0 \, .
\end{equation}
The solution of this equation is given in terms of the Hankel functions of order $1/3$,
\begin{equation}
\label{eq:vH}
    v \propto A_1 \tilde{\phi}^{1/3} H_{1/3}^{(1)}(\tilde{\phi}) + A_2 \tilde{\phi}^{1/3} H_{1/3}^{(2)}(\tilde{\phi}) \,,
\end{equation}
valid in the interval between $y_1-\delta_E$ and $y_1$. In the vicinity of $y_1-\delta_E$, the same equation (\ref{eq:vxx}) can be solved by the WKB approximation provided 
\begin{equation}
\tilde{\phi}(y_1-\delta_E) \gg 1 \,.    
\end{equation}
In this case the solution is a linear combination of the two solutions
\begin{equation}
v_{\mathrm{WKB}\pm} \propto \frac{1}{\sqrt{K(y)}} \exp (\pm i \tilde{\phi})    \, .
\end{equation}
This WKB approximation can be extended to all the regions where the conditions
\begin{align}
&\left|\frac{K''}{K'} \right| \ll K  \, ,   \\
&\left|\frac{K'}{K} \right| \ll K \, .
\end{align}
are satisfied and in particular initially around $y\to -\infty$ where 
the normalisation of the solution can be determined by imposing that the Wronskian should be
\begin{equation}
W= v v'^*-v^*v' = i \,. 
\end{equation}
This condition comes from imposing the canonical quantisation condition $[v(\vec x,\eta), \pi_v (\vec y, \eta)]=i \delta^{(3)}(\vec x-\vec y)$ between the field $v$ and its conjugate momentum at equal time. This yields
\begin{equation}
v_{\mathrm{WKB}} = \frac{1}{\sqrt{2{\tilde k}}} \frac{1}{\sqrt{K(y)}} \exp (i \tilde{\phi})    \, .
\end{equation}
We match this with Eq. (\ref{eq:vH}) around $y_1-\delta_E$ where $\phi(y_1-\delta_E)\gg 1 $ and using the asymptotic behaviour of the Hankel functions to find
\begin{equation}
\label{eq:vsol}
    v(y) = \frac{\sqrt{\pi}}{2 \sqrt{{\tilde k}}}\ \left(\frac{2}{-3 s_1}\right)^{1/6}  \exp\left(\frac{5\pi}{12}i\right)\tilde{\phi}^{1/3}(y) H_{1/3}^{(1)}(\tilde{\phi}(y)) \, ,
\end{equation}
where $s_1=  \left[ \left.\frac{d (K^2(y))}{dy}\right|_{y=y_1}\right] $.

We can now  evaluate the power spectrum  at the time of horizon crossing, $y_H=-1 \approx y_1 = {\tilde k} \eta_1$. We have to evaluate Eq. (\ref{eq:vsol}) in the limit $y \to y_1$ corresponding to $\tilde{\phi} \to 0$,
\begin{equation}
\label{eq:v2}
    |v(y_1)|^2 \to \frac{1}{2 {\tilde k} \pi }\ \left(\frac{1}{3 |s_1|}\right)^{1/3} \Gamma(1/3)^2 \,.
\end{equation}

The power spectrum is then given by
\begin{align}
\label{eq:PSkx1}
\mathcal{P}_{\zeta}({\tilde k},y_1)  & =\frac{{\tilde k}^3}{2 \pi^2} \left|\frac{v(y_1)}{z(\eta_1)}\right|^2 \nonumber \\
& = \frac{y_1^2}{4 \pi^3} \frac{  \left(\frac{1}{3 s_1}\right)^{1/3} \Gamma(1/3)^2 }{\frac{1}{h_{\mathrm{dS}}^2 } f_2 {\cal A} \bigg( 1 + \frac{\alpha }{2f_2} \Big( \frac{{\cal A}_{1}}{\bar{\cal A}} 
+ \frac{1}{h_{\mathrm{dS}}^2 y_1^2 }\frac{1}{1+\alpha  \frac{27 \beta_K }{8 f_2 (1+\alpha_B)^2}} \frac{{\cal A}_{2}}{\bar{\cal A}} \Big) \bigg)} \,.
\end{align}
Hence, $\mathcal{P}_{\zeta}$ is constant and the spectral index $n_s$ is given by
\begin{align}
n_s - 1 &  = 0 \, .
\end{align}
As a result, the spectrum is scale invariant and cannot be used as a sound inflationary model. The scale invariance of the spectrum results from the fact that the Lagrangian of derivative DHOST models with derivative scordurata perturbations does not involve $\phi$, only its derivatives. More precisely a rescaling of $\eta \to \lambda \eta$ appears as a shift of time $\tilde t\to \tilde t -\frac{1}{h_{\rm dS}}\ln \lambda $ which can be absorbed by a shift $\varphi\to \varphi -\frac{1}{h_{\rm dS}}\ln \lambda$ in a shift symmetric theory. The rescaling $\eta\to \lambda \eta$ corresponds to the change $k\to \lambda^{-1} k$ at horizon crossing. The invariance under the shift symmetry guarantees the scale invariance of the spectrum. We will obtain a breaking of scale invariance by introducing polynomial interactions below \cite{ArkaniHamed:2003uz}.

\begin{figure}[!h]
\centering
\includegraphics[width=0.75\linewidth]{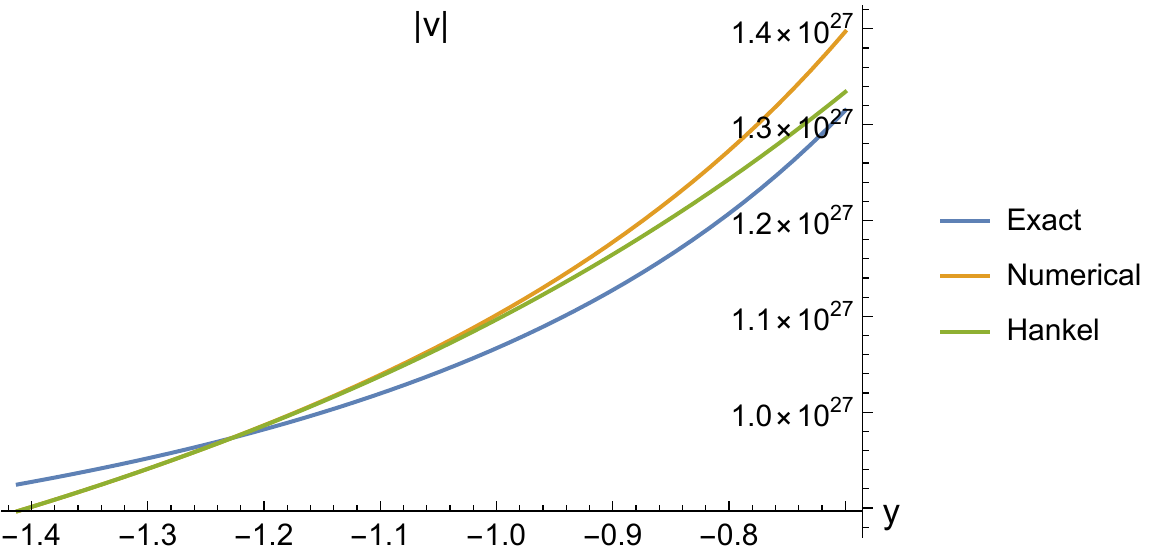} 
\caption{The modulus of the Mukhanov-Sasaki variable, evaluated at the pivot scale $\tilde{k}_\star$,  $|v(\tilde{k}_*,(y)|$, for the exact (blue), numerical (orange) and the WKB improved (Hankel) (green) solutions obtained in the case $\alpha=0$. At horizon crossing, the accuracy of the numerical/ WKB improved solutions is at the 3\% level.}
\label{fig:absv}
\end{figure}

In order to confirm  our results, we compare the exact analytical result (Eq. \ref{DHOST:scalar}) with the improved WKB one (Eq. \ref{eq:vsol}) for the $\alpha=0$ scenario. Additionally, we solve the equation for $v$ numerically from $y_1$ up to $y_H=-1$ using the initial conditions given by the limit $y \to y_1$ (Fig. \ref{fig:absv}). The plots show the very good agreement between the three curves at $\tilde{k}=\tilde{k}_*$.

\begin{figure}[!h]
\centering
\includegraphics[width=0.75\linewidth]{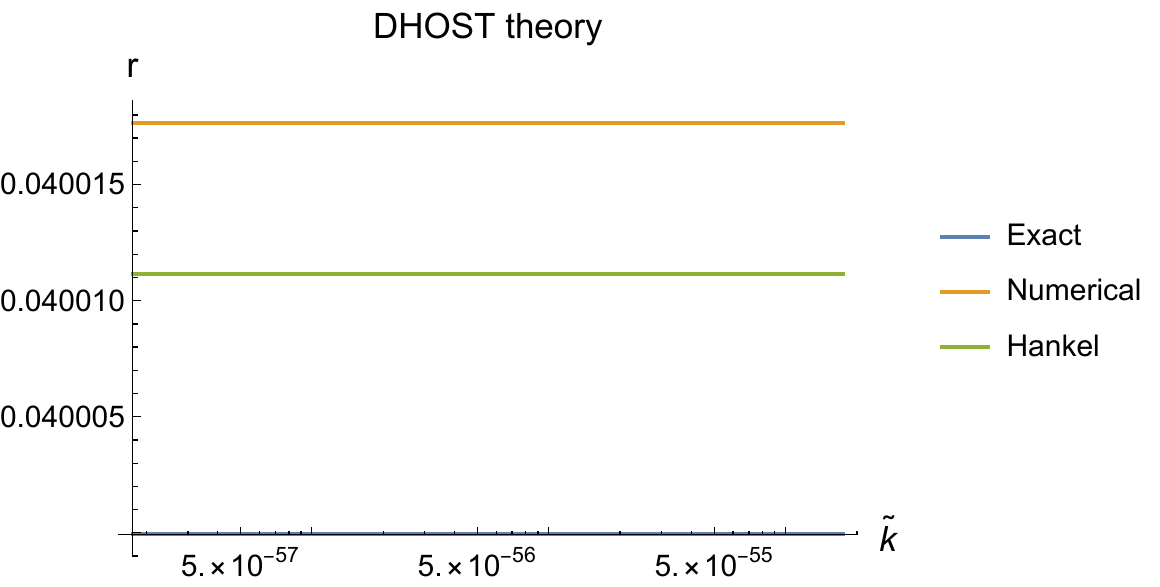} 
\caption{Tensor-to-scalar ratio in DHOST theories ($\alpha_B=1$, $\alpha_H=1.04$ and $\beta_K=3.97343$), showing the agreement between the exact solutions and the WKB one. The analytical solution (blue) is that of Eq. (\ref{r:DHOST}), the matched WKB Hankel solution (green) is obtained by using equations analogous to Eq. (\ref{eq:vsol}), while the numerical solution (orange) is obtained by numerically integrating for $v$ from $y_1$ to $y_H$. The three curves are in good agreement at a level of 0.7\%.}
\label{fig:rDHOST}
\end{figure}

\section{Shift-symmetry breaking perturbations}
\label{sec:cos}
\subsection{The models and their spectrum}
The scordatura models presented in the previous section all rely on derivative corrections to the stealth DHOST solutions, and as the DHOST solutions themselves, predict scale invariant power spectra. In this section, we show that by considering shift-symmetry breaking  perturbations around a DHOST background we can break the exact scale invariance of the perturbations and hence  determine a spectrum that matches the observed density perturbations. The breaking of scale invariance by such small polynomial interactions is certainly generic from an effective field theory point of view as one can expect that the original shift symmetry $\phi\to \phi+c$ of the DHOST models will be broken, for instance by non-perturbative effects like in the case of the axions. 

Hence, we start with a DHOST action (\ref{action-dhost}) as the background and we consider perturbations as 
\begin{equation}
\label{eq:lambdam2}
   S_{\rm V} = -\int d^4  x \sqrt{- g}  \mu^4 \left(\cos \frac{\phi}{f}-1\right) \,.
\end{equation}
When $\phi \ll f$ we have the expansion
\begin{equation}
S_{\rm V} = \int d^4 x \sqrt{-g} \bigg[ - \frac{m_{\rm phys}^2}{2}  \phi^2 - \frac{\lambda_{\rm phys}}{4!}  \phi^4 \bigg] \,,    
\end{equation}
where $m^2_{\rm phys}<0$  and 
\begin{equation}
m^2_{\rm phys}=- \frac{\mu^4}{f^2},\ \lambda_{\rm phys}= \frac{\mu^4}{f^4} \, .
\end{equation}
This can be written in reduced units as
\begin{equation}
S_{\rm V} = \int d^4 \tilde x \sqrt{-\tilde g} \bigg[ - \frac{m_{}^2}{2}  \varphi^2 - \frac{\lambda_{}}{4!}  \varphi^4 \bigg] \,,    
\end{equation}
where $g_{\mu\nu}= \Lambda^2 \tilde g_{\mu\nu}$ and $\tilde g_{\mu\nu}$ is the dimensional metric depending on $\eta$ and $\tilde x$.
In terms of the  numerical parameters $m^2$ and $\lambda$ we have
\begin{equation}
    f= \frac{\sqrt{\vert m^2 \vert}}{\sqrt \lambda} M, \ \mu= \frac{\sqrt{\vert m^2 \vert}}{ \lambda^{1/4}} \Lambda  \, .
\end{equation}
We must also require that $\phi\lesssim  f$ to use the quartic expansion. This is valid when
\begin{equation}
\varphi \lesssim \frac{\sqrt{\vert m^2 \vert}}{\sqrt \lambda} \,.
\label{crucial}
\end{equation}
We must also impose that the perturbation is small compared to the energy density of inflation.

We proceed similarly to the previous sections, obtaining the second-order perturbed action of the same form as Eq. (\ref{Lagrangian-scordatura}), but with coefficients
\begin{align}\label{matrices-components-m2l4}
& {\bar k}_{11} = 0 , \hspace{2.5cm} 
{\bar k}_{12} = 0  , \hspace{2.5cm} 
{\bar k}_{22} = 0  , \\ \nonumber 
& {\bar n}_{12} = 0 , \hspace{2.5cm} 
{\bar n}_{13} =  0 , \hspace{2.5cm} 
{\bar n}_{23} = 0 , \\ \nonumber 
& {\bar m}_{11}  =  \frac{1}{8} a^3 (12 m^2 \varphi^2 + \lambda \varphi^4) , 
\\ \nonumber 
& {\bar m}_{12} =  -\frac{1}{8} a^3 (-1+ \alpha_H) (12 m^2 \varphi^2 + \lambda \varphi^4)  ,  \nonumber \\ 
& {\bar m}_{22} = \frac{1}{24} a^3 (-1-6 \alpha_H+3 \alpha_H^2) (12 m^2 \varphi^2 + \lambda \varphi^4) , \nonumber \\
& {\bar m}_{22{\rm s}} = 0 \,, \hspace{1cm} {\bar m}_{23} = 0 \,, \hspace{1cm} 
{\bar m}_{33{\rm s}} = 0 \,, \nonumber \\ \nonumber 
& {\bar m}_{33} = -\frac{1}{24} a (12 m^2 \varphi^2 + \lambda \varphi^4) \,. 
\end{align}
As before, we take linear de Sitter solutions, but of the form
\begin{equation}
\varphi({\tilde t}) = c-{\tilde t}\, ,
\end{equation}
replacing Eq. (\ref{stealth-sol}), where $c$ is a constant corresponding to the initial value of the field. We also make sure that the energy density due to the potential term  is a small perturbation to the background energy density which drives inflation. In particular we have the correction to the Friedmann equation due to the potential term (\ref{eq:lambdam2}), and the Einstein (\ref{Stealth-Raychuadhuri}-\ref{Stealth-Friedmann}) are modified to
\begin{align}
& f_0 +  6 h_{\rm dS}^2  f_2 + \tilde V(\varphi)  = 0 \,, \label{Raychuadhuri2}
\\
& f_{0,{\mathrm x}} + 3 h_{\rm dS} ( 4 h_{\rm dS} f_{2,{\mathrm x}} + \mu f_{1,{\mathrm x}} ) = 0 \,, \label{Friedmann2}
\end{align}
where we have defined $\tilde V(\varphi)= \Lambda^{-4} V(M \varphi)$. The de Sitter background is preserved if $\tilde V$ is negligible.

Using the definition of the Hubble parameter and Eq. (\ref{eq:a}), we switch from cosmic time to conformal time with
$
{\tilde t}= -\frac{1}{h_{\mathrm{dS}}} \log(-h_{\mathrm{dS}} \eta)   \, , 
$
valid during the de Sitter phase. 
We then follow the same steps as before for the quantisation, but we keep the full $m^2$ and $\lambda$ dependence. 
The terms involving the perturbations  feature logarithmic terms that break scale-invariance. Hence, $v$ satisfies
\begin{equation}
v''+K^2(y,k) v=0 \, ,    
\end{equation}
where the primes denote derivatives with respect to $y$. The expression for $K^2(y)$ takes the following form now 
\begin{align}
    &K^2({\tilde k},y)=  -\frac{z''}{z}\frac{1}{{\tilde k}^2} -\frac{{\tilde k}}{48 \left(\frac{6 f {\tilde k} \left(\alpha_B^2+2 \alpha_B-\beta_K+1\right)}{(\alpha_B+1)^2}-\frac{d {\tilde k} \left((\alpha_B+1)^2 \left(3 \alpha_H^2-6 \alpha_H-1\right)+\frac{9 {\tilde k}^4 \left(\alpha_B^2+2 \alpha_B-\beta_K+1\right)^2}{y^2 \left({\tilde k}^2-\frac{\beta_K d {\tilde k}^2 }{32 (\alpha_B+1)^2 f h_{\mathrm{ds}}^2 y^2}\right)^2}\right)}{48 (\alpha_B+1)^4 h_{\mathrm{ds}}^2}\right)} \nonumber \\
    &\times \left\{
    \frac{96 f (\alpha_B-\alpha_H)}{\alpha_B+1} + \frac{c+\frac{\log \left(-\frac{h_{\mathrm{ds}} y}{{\tilde k}}\right)}{h_{\mathrm{ds}}}}{h_{\mathrm{ds}}^2} \right. \nonumber \\
    & \times  \left[\left(-\frac{(\alpha_H+1)^2}{(\alpha_B+1)^2}-\frac{3}{y^2}\right) \left(c+\frac{\log \left(-\frac{h_{\mathrm{ds}} y}{{\tilde k}}\right)}{h_{\mathrm{ds}}}\right) \left(\lambda  \left(c+\frac{\log \left(-\frac{h_{\mathrm{ds}} y}{{\tilde k}}\right)}{h_{\mathrm{ds}}}\right)^2+12 m^2\right) \right. \nonumber \\
    &  
    + \left( 3072 (\alpha_B+1) f^2 h_{\mathrm{ds}} y^2 \left(2 \alpha_B^2+4 \alpha_B-(\alpha_H+1) \beta_K+2\right)  \left(h_{\mathrm{ds}}^2 \left(c^2 \lambda  (3 c h_{\mathrm{ds}}-4)+12 m^2 (3 c h_{\mathrm{ds}}-2)\right) \right. \right. \nonumber \\
    &  \left. \left.  +h_{\mathrm{ds}} \left(9 h_{\mathrm{ds}} \left(c^2 \lambda +4 m^2\right)-8 c \lambda \right) \log \left(-\frac{h_{\mathrm{ds}} y}{{\tilde k}}\right)+\lambda  (9 c h_{\mathrm{ds}}-4) \log ^2\left(-\frac{h_{\mathrm{ds}} y}{{\tilde k}}\right)+3 \lambda  \log ^3\left(-\frac{h_{\mathrm{ds}} y}{{\tilde k}}\right)\right) \right)/  \nonumber \\
    & \left. \left. (-32 f_2 h_{\mathrm{ds}}^2 y^2 (1 + \alpha_B)^2 + d  \beta_K)^2
    \right]
    \right\} 
\end{align}
with
\begin{align}    
    &z^2 = \frac{f_2 {\tilde k}^2}{h_{\mathrm{ds}}^2 y^2} \left\{ \frac{6 \left(\alpha_B^2+2 \alpha_B-\beta_K+1\right)}{(\alpha_B+1)^2}- \frac{d}{48 f_2 h_{\mathrm{ds}}^2 (1 + \alpha_B)^4} \right. \nonumber \\
    & \left. \quad \times \left( (\alpha_B+1)^2 \left(3 \alpha_H^2-6 \alpha_H-1\right)+\frac{9 {\tilde k}^4 \left(\alpha_B^2+2 \alpha_B-\beta_K+1\right)^2}{y^2 \left(k^2-\frac{\beta_K {\tilde k}^2  d}{32 (\alpha_B+1)^2 f h_{\mathrm{ds}}^2 y^2}\right)^2} \right) \right\}
\label{eq:vm2l}    
\end{align}
where
\begin{equation}
d \equiv \left(c+\frac{\log \left(-\frac{h_{\mathrm{ds}} y}{{\tilde k}}\right)}{h_{\mathrm{ds}}}\right)^2 \left(\lambda \left(c+\frac{\log \left(-\frac{h_{\mathrm{ds}} y}{{\tilde k}}\right)}{h_{\mathrm{ds}}}\right)^2+12 m^2\right) \, .
\end{equation}
As $m^2 \ll 1$ and $\lambda \ll 1$, the turning point where  $\tilde{K}^2(y)=0$ at  $y_1(k)$ still exists.  We will determine it numerically below. The value of the derivative of $\tilde{K}^2(y)$ at the turning point for values of the parameters that are observationally interesting  will also be determined numerically, allowing one to use Eq. (\ref{eq:v2}).
The power spectrum is determined using the asymptotic matching procedure outlined in the scordurata case and compared to 
the numerical results
in section \ref{sec:num}, by performing a numerical integration from $y_1$ to $y_H=-1$ using the asymptotic form of Eq. (\ref{eq:v2}) at $y_1$ as initial conditions.  
The scalar power spectrum becomes
\begin{align}
\label{PSm2l}
\mathcal{P}_{\zeta} ({\tilde k},y_H)
& = \frac{{\tilde k}^3}{2 \pi^2} \left| \frac{v({\tilde k},y_H,m^2,\lambda)}{z({\tilde k},y_H,m^2,\lambda)}\right|^2  \, ,
\end{align}
where $y_H$ corresponds to the horizon position $y_H=-1$.
The spectral index and the running of the spectral index take the forms:
\begin{align}
n_s({\tilde k},y_H) & = 1 + \frac{d \log(\mathcal{P}_{\zeta}({\tilde k},y_H))}{d \log({\tilde k})}  \, , \\
\alpha_s ({\tilde k},y_H) & = \frac{ d n_s({\tilde k},y_H)}{d \log({\tilde k})} \, , \\
\beta_s ({\tilde k},y_H) & = \frac{ d \alpha_s({\tilde k},y_H)}{d \log({\tilde k})} \, .
\end{align}
This provides enough details to compare with the Planck data. We will also require the gravitational wave spectrum of the theories.

\subsection{Tensor perturbations}

We  determine the spectrum of tensor perturbations in the DHOST and perturbed models. The second order action of tensor modes of a DHOST action (\ref{action-dhost}) with perturbations (\ref{eq:lambdam2}) can be written in conformal time as
\begin{align}
S_2^{\mathrm{tensor}}=\int d\eta \slashed{d}^3 k \left[a^2 f_ 2E_{ij}'E^{ij \prime}  - a^2 f_2 k^2 E_{ij}E^{ij} + \frac{1}{24} E_{ij}E^{ij} a^4 (12 m^2 \varphi^2 + \lambda \varphi^4)  \right] \, .
\end{align}
Writing $\mu_T = z_T E$, where $z_T^2=a^2 f_2$, and taking the Euler-Lagrange equations, $\mu_T$ satisfies the following equation
\begin{equation}
\mu_T''+\left[{\tilde k}^2 -\frac{1}{24 f_2} \frac{1}{h_{\mathrm{ds}}^2 \eta^2}\left( 12 m^2 (c+\frac{1}{h_{\mathrm{ds}}} \log(-h_{\mathrm{ds}} \eta))^2 + \lambda (c+\frac{1}{h_{\mathrm{ds}}} \log(-h_{\mathrm{ds}} \eta))^4 \right) - \frac{2}{\eta^2} \right] \mu_T = 0    \, .
\end{equation}
In the DHOST case ($m^2=\lambda=0$), when taking into account the two polarizations of the graviton, we find
\begin{align}
\mu_T({\tilde k},y)&=\frac{1}{\sqrt{2 {\tilde k}}}\left(1-\frac{i}{y} \right) \exp(-i y) \,, \\
z_T({\tilde k},y)&=\frac{{\tilde k}^2 f_2}{h_{\mathrm{ds}}^2 y^2} \,, \\
\label{DHOST:tensor}
\mathcal{P}_T({\tilde k},y)&=\frac{{\tilde k}^3}{2 \pi^2} \left|\frac{\mu_T}{z_T} \right|^2 = \frac{h_{\mathrm{ds}}^2 y^2 \left(1+\frac{1}{y^2} \right)}{4 \pi^2 f_2} \,.
\end{align}
Hence, using Eq. (\ref{DHOST:scalar}) we can calculate the tensor-to-scalar ratio $r$,
\begin{equation}
r_{\mathrm{DHOST}}({\tilde k},y)= \frac{\mathcal{P}_T({\tilde k},y)}{\mathcal{P}_{\zeta}({\tilde k},y)}= 6 \bar{c}_s \frac{1+\frac{1}{y^2}}{1+\frac{1}{\bar{c}_s^2 y^2}} \left(1-\frac{\beta_K}{(1+\alpha_B)^2}\right) \, ,
\end{equation}
which is constant. We fix the horizon to $y_H=-1$, to obtain
\begin{equation}
\label{r:DHOST}
r_{\mathrm{DHOST}}=   \frac{12 \bar{c}_s \left(1-\frac{\beta_K}{(1+\alpha_B)^2}\right)}{1+\frac{1}{\bar{c}_s^2}}  \, .
\end{equation}
In the following, we shall impose that the tensor to scalar ration of DHOST models satisfies the \textit{Planck} constraints and then perturb by the polynomial interactions to see how this ratio evolves in $k$, in particular we shall evaluate it at the pivot scale $k_\star$ used by the \textit{Planck} collaboration.

\subsection{Physical scales and constraints from data}

Next we investigate the values of the reduced parameters that are of observational interest by switching back to the dimensional parameters. We will choose the remaining parameters such that the \textit{Planck} inflationary constraints are satisfied.

The \textit{Planck} 2018 release has determined the tightest constraints on the spectral index and its derivatives. By considering a pivot scale of $k_* = 0.05 \mathrm{~Mpc}^{-1}$, these are given by
\begin{align}
\label{ns_const}
n_s &= 0.9625 \pm 0.0048    \, ,\\
\alpha_s &= 0.002 \pm 0.010 \, , \\
\beta_s & = 0.010 \pm 0.013 \, , \\
\ln(10^{10} A_s) & = 3.044 \pm 0.014 \,
\label{bs_const}
\end{align}
at $68\%$ confidence level, using the TT,TE,EE+lowE+lensing likelihoods \cite{Akrami:2018odb}.
The observable scales today correspond to $10^{-4}~\mathrm{Mpc}^{-1} \lesssim k \lesssim 10^{-1}~ \mathrm{Mpc}^{-1}$, i.e. perturbations that have reentered the Hubble radius around recombination.
Using the rescaling described in Eq. (\ref{coordinets-ch}), $k$ is such that
$
\tilde{k}=\Lambda^{-1} k
$
In particular the pivot scale becomes $\tilde k_\star = \Lambda^{-1} k_\star$.

Data from CMB experiments have been used to place stringent constraints on the tensor to scalar ratio. The tightest constraints yield $r < 0.044$ at  95 \% confidence \cite{Tristram:2020wbi} when combining \textit{Planck} with BICEP2/Keck 2015 data \cite{Ade:2018gkx}. Future CMB experiments, such as LiteBIRD \cite{Hazumi:2019lys}  will be able to  tighten this bound significantly, up to $r < 10^{-3}$. In the next section, we will choose the parameters appearing in the DHOST model for $r$ (Eq. \ref{r:DHOST}) in order to satisfy these bounds. 

We consider models with perturbations in $m^2$ and  $\lambda$ separately and with both types of perturbations.
We look for models with $h_{\mathrm{ds}} \ll 1$.
We fix the value of $\Lambda$ at the Planck  scale and 
 consider a $\tilde{k}$-range adjusted such that $\tilde{k}_*= 2.62 \times 10^{-59}/h_{\mathrm{ds}}$.

\subsection{Numerical results}
\label{sec:num}
\begin{figure}[!h]
\centering
\includegraphics[width=0.49\linewidth]{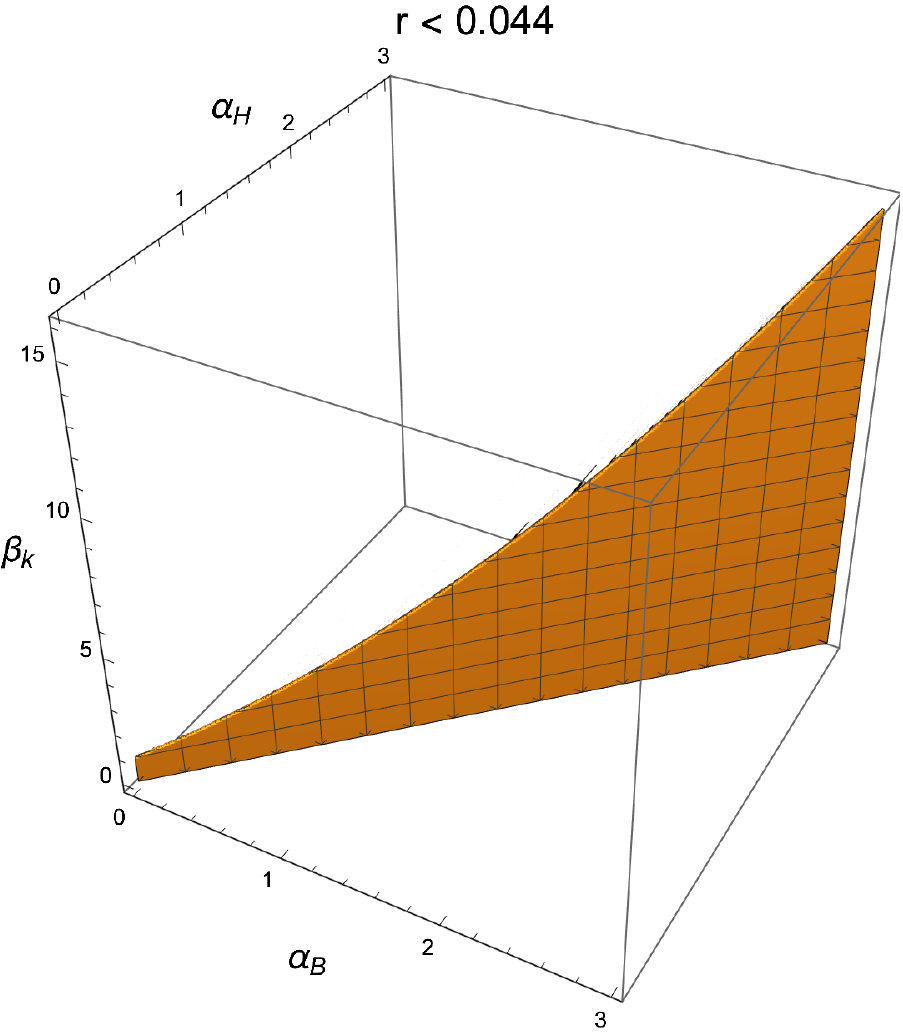} 
\includegraphics[width=0.49\linewidth]{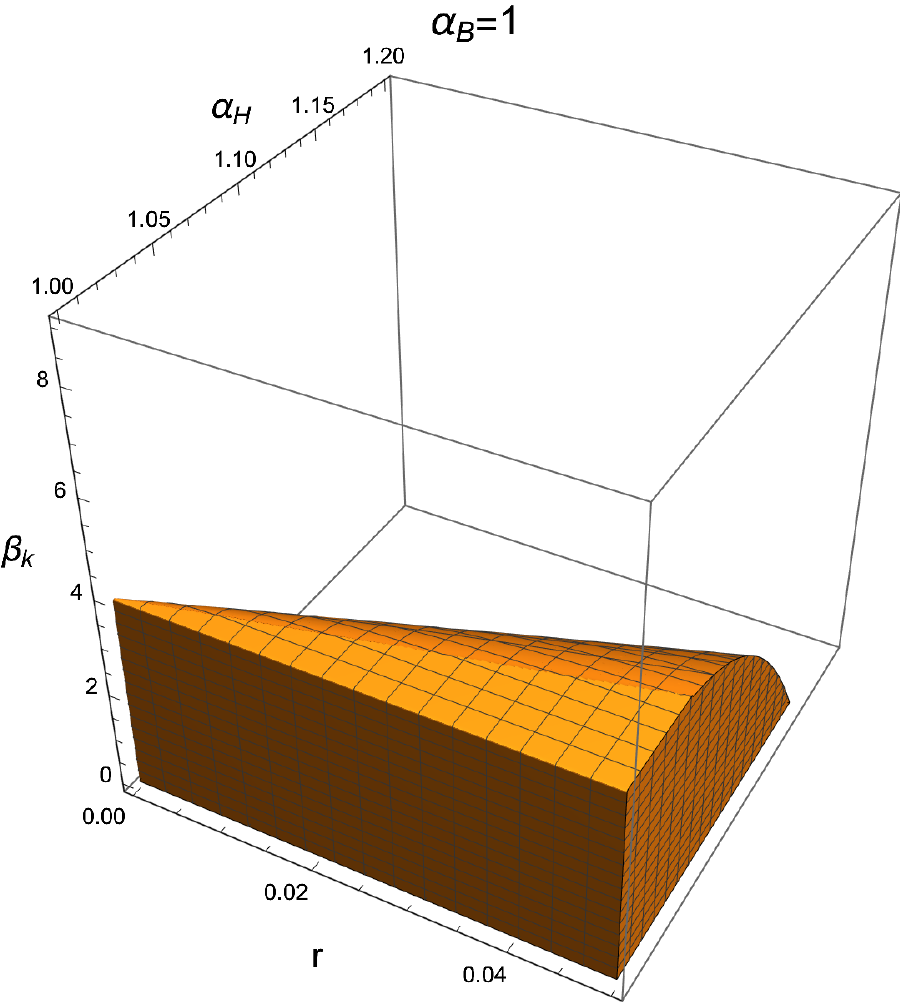} 
\caption{Three-dimensional plot of the DHOST parameter regions where the \textit{Planck}   constraints are satisfied (left) and a three-dimensional plot for the DHOST parameters $\alpha_H$ and $\beta_K$ that yield a tensor-to-scalar ratio less than $r$, when $\alpha_B=1$ (right). {From the left figure we see that the allowed region is very narrow in the $\alpha_B-\alpha_H$ plane, while in the right one, by fixing $\alpha_B=1$, we show how the allowed region shrinks as $r$ is decreased. We also note that $\beta_K$ plays only a minor role, with all values $\beta_K \le (\alpha_B+1)^2$ being in the allowed region.}}
\label{fig:rregions}
\end{figure}

Eq. (\ref{r:DHOST}) can be used to investigate the range of the DHOST parameters such that the constraints on the tensor-to-scalar ratio are satisfied before looking at the perturbations themselves. In Figure \ref{fig:rregions}, we show the parameter space where the \textit{Planck} CMB constraints are satisfied (left). The plots show that the the space is almost degenerate in the $\alpha_B-\alpha_H$ plane, with the allowed region narrowing as $r$ decreases (right).
As in DHOST models the power spectrum is scale-invariant, this cannot be used to fix template values  for $n_s$, but we will show that we can use the $m^2$ and $\lambda$ perturbations to get a spectral index of the correct magnitude at $k_*$. 

In order to illustrate this numerically, we start with a baseline model with $\alpha_B=1$, $\alpha_H=1.04$ and $\beta_K=3.97343$, yielding a DHOST tensor-to-scalar ratio of $r_{\mathrm{DHOST}}=0.04$ and $\bar{c}_s=1.002$. 
Notice that the speed of sound is very close to unity and not problematic. 
 These parameters can be used to determine the derivatives of the functions $f_i$ at $\rm{x}=-1$ using Eqs. (\ref{alpha-i}) and (\ref{beta-K}),
\begin{align}
&f_{2,\rm{x}}= 2.81, \qquad f_{1,\rm{x}} = -6.48 \times 10^{-6}, \qquad f_{0,\rm{x}} = -2.97 \times 10^{-8} \nonumber \\ 
&f_{2,\rm{xx}}= 2.7 \beta_H, \qquad f_{1,\rm{xx}} = -8.1 \times 10^{-5} \beta_B, \qquad f_{0,\rm{xx}}= -2.97\times 10^{-8} \left(\beta_B-4 \beta_H-4.133\right)  \, ,
\end{align}
where $\beta_B$ and $\beta_H$ are free parameters. Notice that the values of the derivatives of the $f_i$ functions vary over many orders of magnitude. We have not performed a thorough scanning of the parameter space, this is left for future work.

\begin{figure}[!h]
\centering
{\large $r=0.04$}\par\medskip
\includegraphics[width = 0.49 \linewidth]{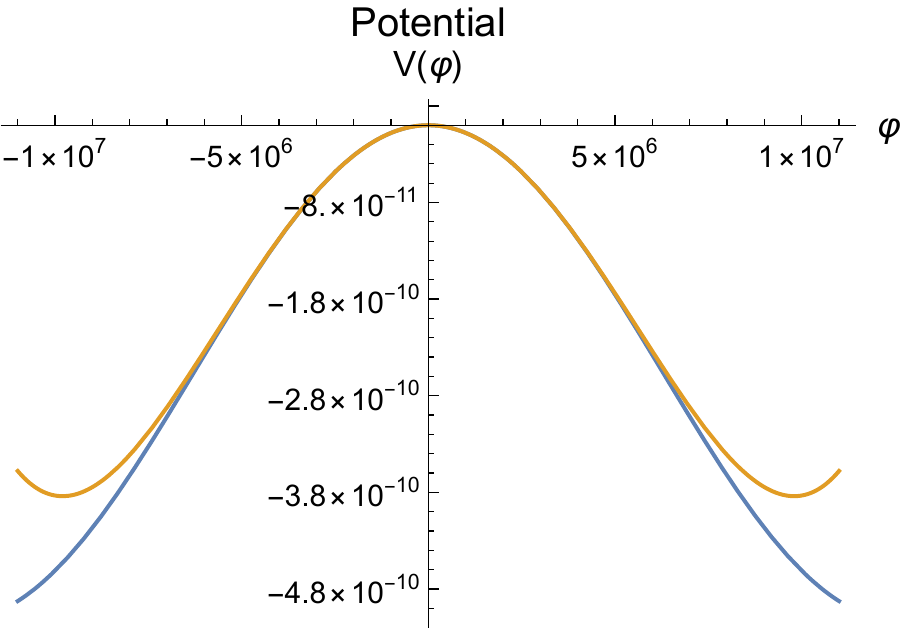} 
\includegraphics[width = 0.49 \linewidth]{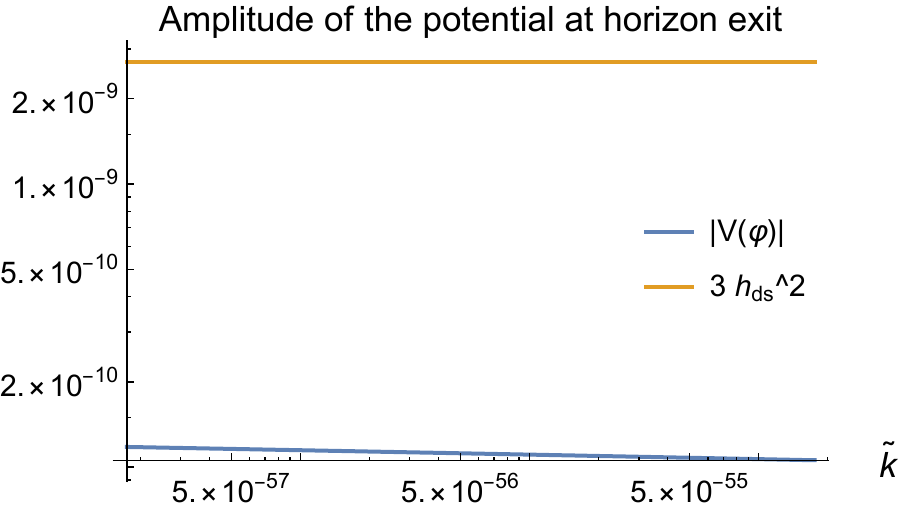} \\
\vspace{8.00mm}
\includegraphics[width = 0.49 \linewidth]{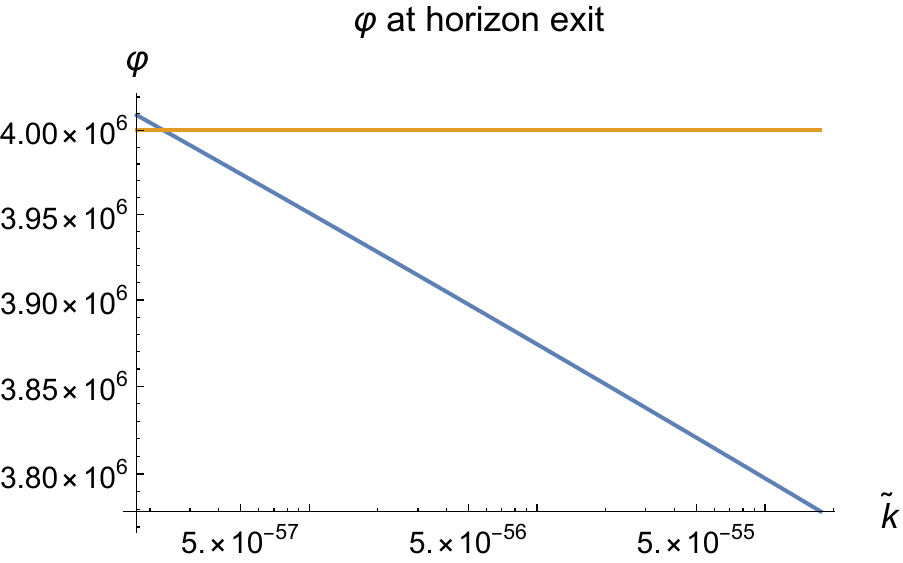} 
\caption{The potential for $\mu/\Lambda = 0.004$ and $f/M=4 \times 10^6$  (blue) and the  corresponding approximation  $m^2 = -1.6 \times 10^{-23}$ and $\lambda =  10^{-36}$ (orange) (left) in the first model. Notice that the amplitude of the potential is negligible compared to the background energy density $3 h_{\rm dS}^2$ (right) at horizon exit.
Notice that the condition (\ref{crucial}) reads $\phi \ll 4 \times 10^6$ and is satisfied here (bottom). The plot shows that at horizon exit the value of the field is close to the origin, and hence the potential is dominated by the $\lambda$ and $m^2$ terms. 
}
\label{fig:rollpot}
\end{figure}

We note that if both perturbation parameters $m^2$ and $\lambda$ are positive, then $r>1$, and  the model are ruled out by current CMB data. We therefore choose a model with $m^2<0$ and $\lambda>0$. This corresponds to the axion-like potentials. Fixing $f_2=2.70$ and $c=1$, the model corresponds to a particle rolling up a potential from $\varphi=1$ at $t=0$, and then upwards towards 0 as illustrated in Fig. \ref{fig:rollpot} (left). In the right panel we show that the amplitude of the potential always remains negligible with respect to the background, while in the bottom panel we show that the condition (\ref{crucial}) is satisfied.
In order to obtain the correct normalisation for the power spectrum at the pivot scale $\tilde{k}_*$,  we fix $h_{\mathrm{ds}}=3 \times 10^{-5}$. This guarantees that the COBE normalisation of the spectrum is satisfied.

We evaluate the scalar and tensor power spectra at horizon crossing $y_H=-1$ using Eq. (\ref{eq:vsol}). In order to check the accuracy of our WKB-improved solutions, we have also integrated numerically from $y_1$ up to $y_H=-1$  for $v$ and compared with the WKB solutions, with the initial conditions taken using the limit (\ref{eq:vsol}) and its derivative as $y \to y_1$.
 Furthermore, for DHOST theories without perturbations, we can also evaluate $r$ exactly (Eq. \ref{r:DHOST}). This is illustrated in Fig. \ref{fig:rDHOST}, where we show that both the numerical and matched WKB  solutions are in excellent agreement with the exact solution at the  0.7\% level.

Numerically, we choose $m^2 = -1.6 \times 10^{-23}$ and $\lambda =  10^{-36}$, such that the spectral index is within the \textit{Planck} limits. By determining its running $\alpha_s$ and the running of its running $\beta_s$ numerically, we show that they are also within the latest inflationary bounds, with $A_s=2.04 \times 10^{-9}$, $n_s = 0.966 $, $\alpha_s = 0.00059$, $\beta_s = 0.000019$ and $r=0.0074$ at the pivot scale. For the cosine potential we get the parameters $f/M=4 \times 10^6$ and $\mu/\Lambda=0.004$, as compatible with a low energy model of inflation. Notice that the scale $M$ is a free parameter. We have chosen $\Lambda=m_{\rm Pl}$ implying that the axion-like perturbation is at a sub-Planckian level. The range of the field is sub-Planckian as long as $f$ is well below the Planck scale itself.

\begin{figure}[!h]
\centering
{\large $r=0.04$}\par\medskip
\includegraphics[width=0.92\linewidth]{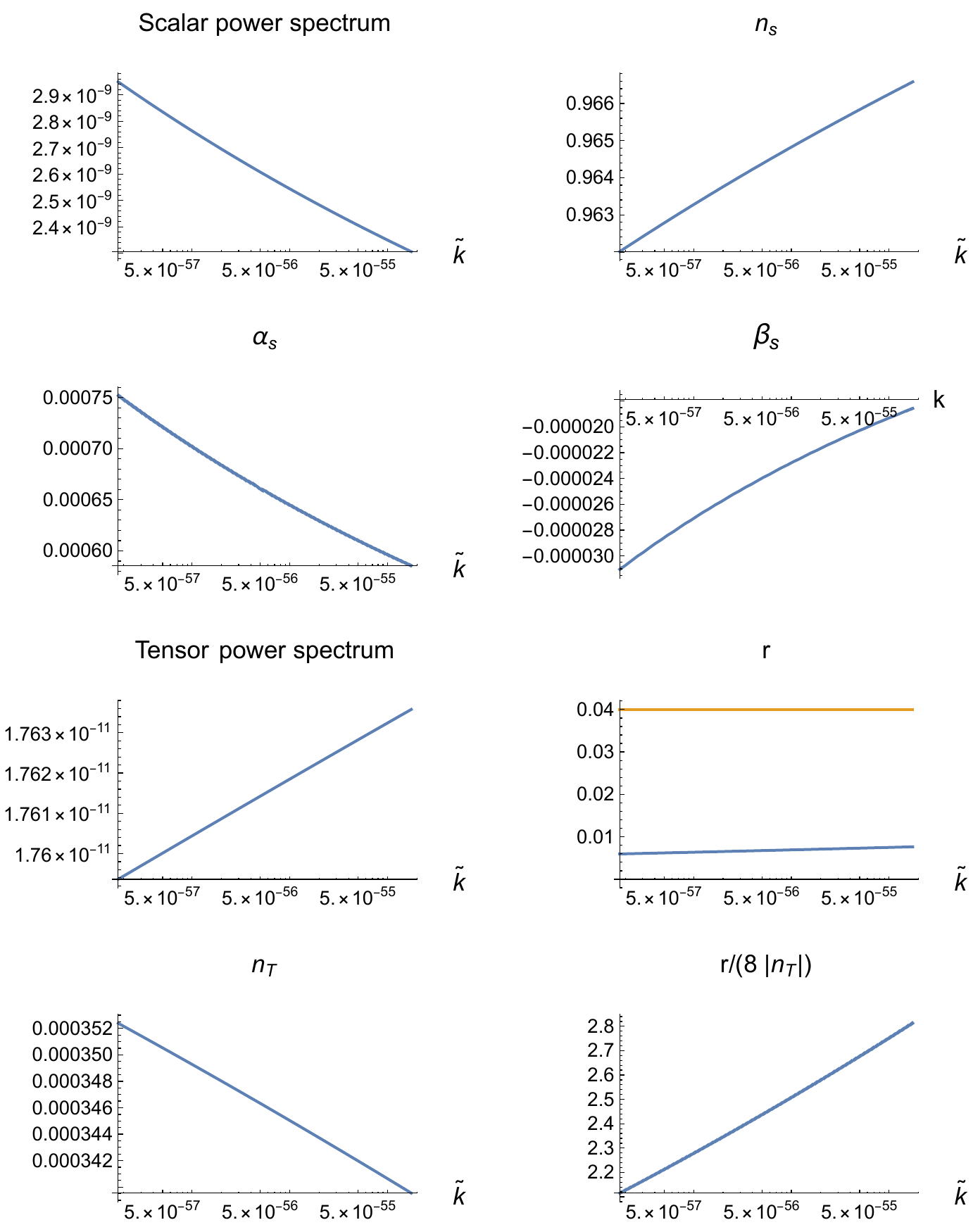} 
\caption{Model compatible with Planck 2018: from top to bottom and left to right, the scalar power spectrum, the scalar spectral index $n_s$, its running $\alpha_s$, the running of its running $\beta_s$, the tensor spectral index, the tensor-to-scalar ratio $r$ (and as a comparison the constant DHOST value), the tensor spectral index and the value of $\frac{r}{8 |n_T|}$ for a DHOST model with $\alpha_B=1$, $\alpha_H=1.04$ and $\beta_K=3.97343$ and perturbations  $m^2 = -1.6 \times 10^{-23}$ and $\lambda =  10^{-36}$. The Hubble constant is $h_{\mathrm{ds}}=3 \times 10^{-5}$.}
\label{fig:allPT}
\end{figure}

In Figure \ref{fig:allPT}, we represent the scalar power spectrum, $n_s$, $\alpha_s$ and $\beta_s$ as a function of $k$, the tensor power spectrum, the tensor-to-scalar ratio, the tensor spectral index $n_T$ and the consistency relation between $r$ and $-8 n_T$. We see that $r$ is smaller than the DHOST value, converging towards it at smaller scales. The consistency relation $r=-8 n_T$ is violated in this model with the ratio $r/(8 n_T)$ increasing towards smaller scales. 
These results show that we can indeed fix the perturbation parameters $m^2$ and $\lambda$ such that the current \textit{Planck} constraints are satisfied and therefore this is a viable class of inflationary models.

\begin{figure}[!h]
\centering
{\large $r=10^{-3}$}\par\medskip
\includegraphics[width=0.92\linewidth]{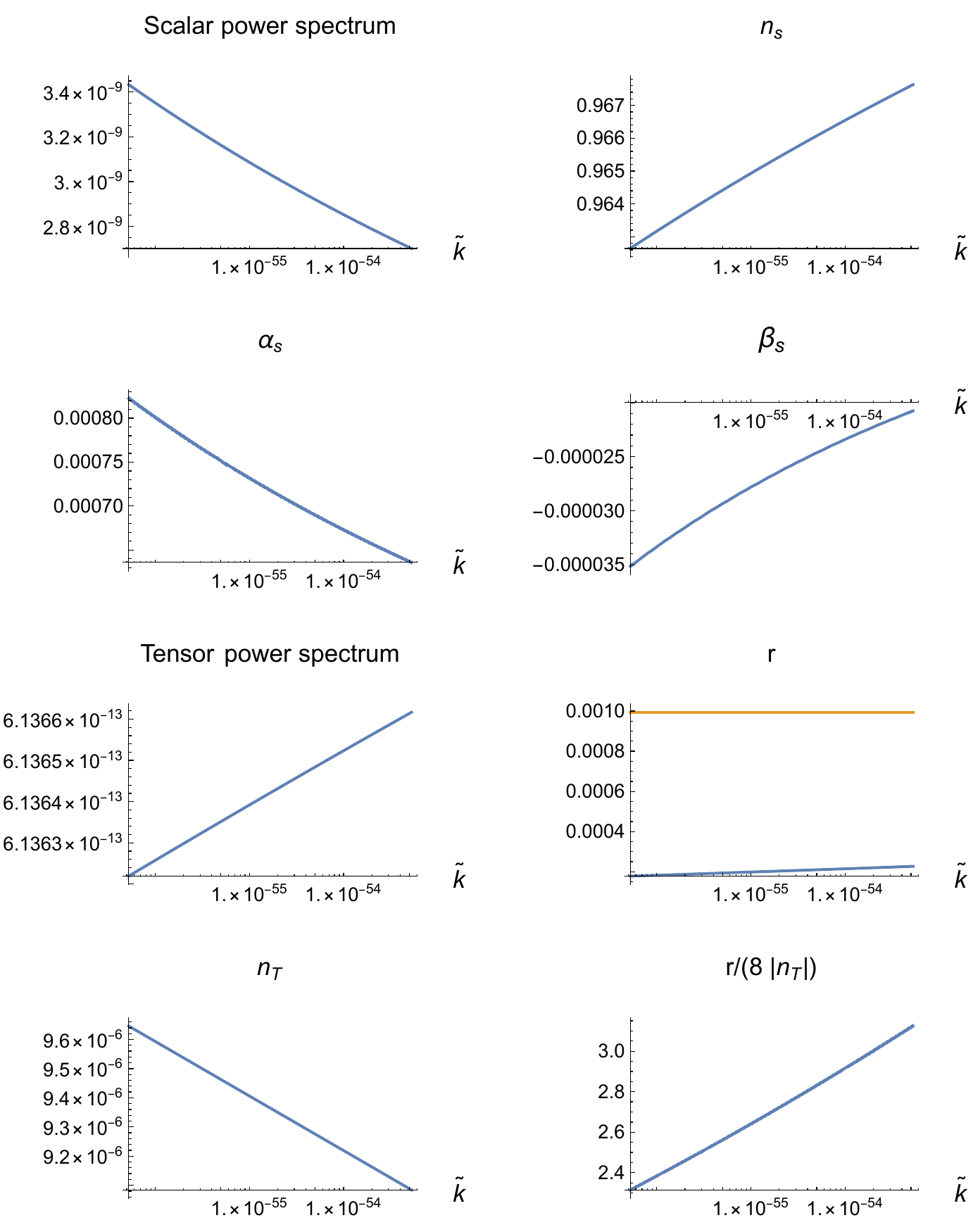} 
\caption{Future experiment values for $r\simeq 10^{-3}$: from top to bottom and left to right, the scalar power spectrum, the scalar spectral index $n_s$, its running $\alpha_s$, the running of its running $\beta_s$, the tensor spectral index, the tensor-to-scalar ratio $r$ (and as a comparison the constant DHOST value), the tensor spectral index and the value of $\frac{r}{8 |n_T|}$ for a DHOST model with $\alpha_B=1$, $\alpha_H=1.001$ and $\beta_K=3.9993$ and perturbations $\lambda= 5 \times 10^{-32}$ and  $m^2= -1.5 \times 10^{-27}$. The Hubble constant is $h_{\mathrm{ds}}= 10^{-5}$.}
\label{fig:allPTsmallr}
\end{figure}

\begin{figure}[!h]
\centering
{\large $r=10^{-3}$}\par\medskip
\includegraphics[width=0.49 \linewidth]{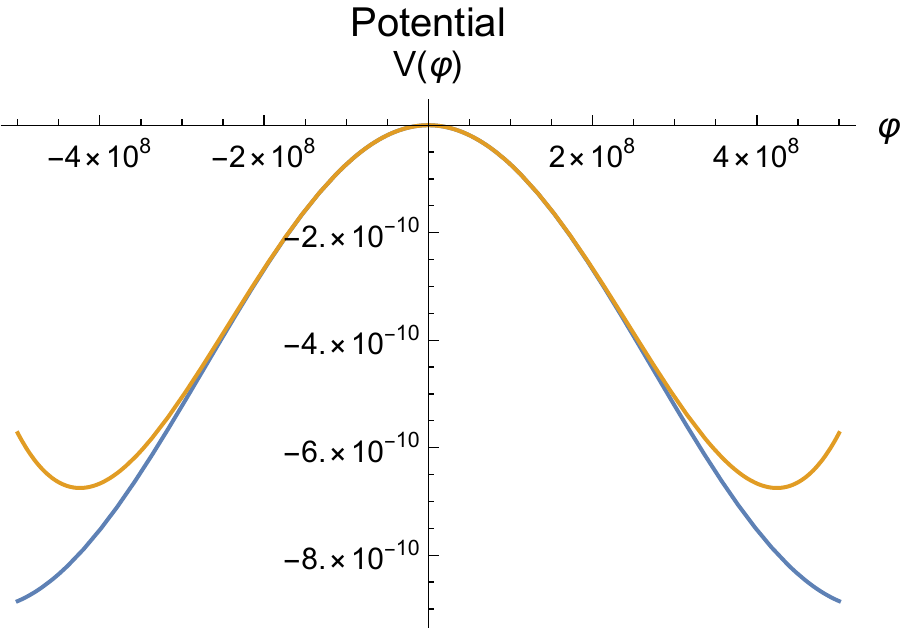} 
\includegraphics[width=0.49 \linewidth]{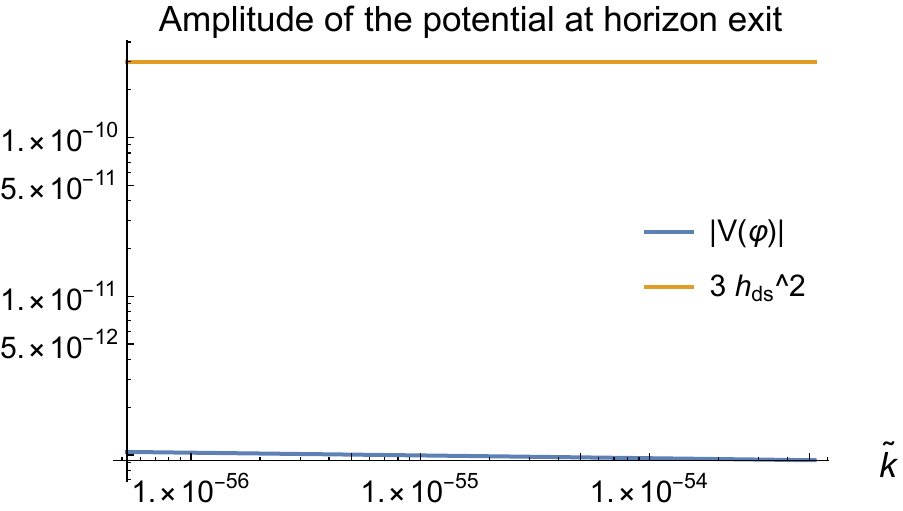} \\
\vspace{8.00mm}
\includegraphics[width=0.49 \linewidth]{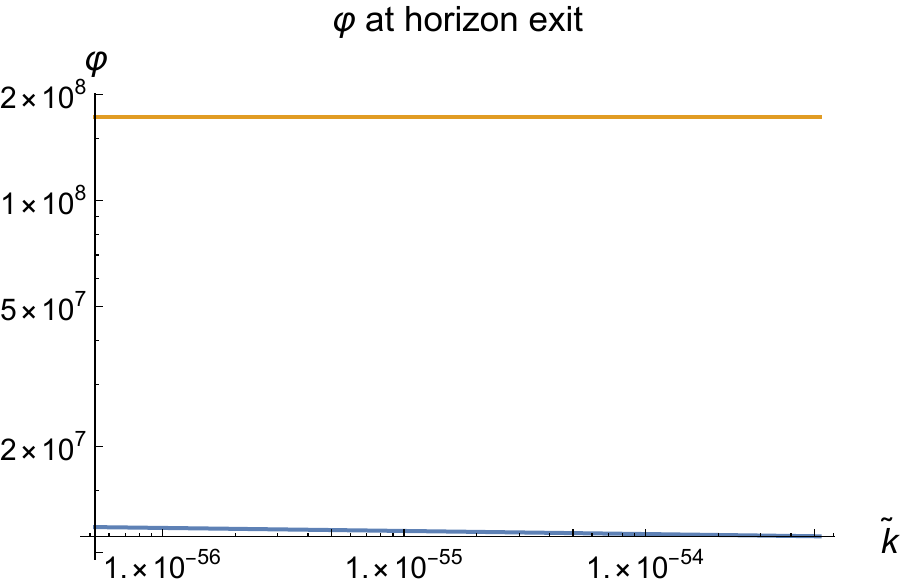} 
\caption{The potential in the second perturbation model for $\mu/\Lambda = 0.0046$ and $f/M=1.73 \times 10^8$ (blue) and the  corresponding approximation  $m^2 = -1.5 \times 10^{-26}$ and $\lambda =  5 \times 10^{-43}$ (orange) (left). The amplitude of the potential is negligible compared to the background energy density $3 h_{\rm dS}^2$ (right) at horizon exit. The field $\varphi$ at horizon exit, showing that the condition (\ref{crucial}) is satisfied (bottom). At horizon exit, the value of the field is close to the origin, and hence the potential is dominated by the $\lambda$ and $m^2$ terms. 
}
\label{fig:rollpotr}
\end{figure}

We have also investigated a model which has a significantly lower tensor-to-scalar ratio, comparable to the one which will be detectable by LiteBIRD \cite{Hazumi:2019lys} and with a scalar spectral index compatible with \textit{Planck}. In order to find parameters that are are compatible with these constraints, we have first fixed the DHOST parameters $\alpha_B$, $\alpha_H$ and $\beta_K$ to get the relevant tensor-to-scalar ratio, and then we concentrated on the perturbations $m^2$ and $\lambda$ to fix $n_s$. When fixing the DHOST parameters, we observed that the parameter space becomes quite narrow in the the $\alpha_B-\alpha_H$ plane (Fig. \ref{fig:rDHOST}), and that in order to be able to maintain  the accuracy of the numerical computations we also required that $\bar{c}_s = \mathcal{O}(1)$. Hence, we require $\alpha_H \gtrsim \alpha_B$ and $\beta_K \lesssim (\alpha_B+1)^2$. For the perturbations we similarly observe that the scalar spectral index is sensitive to the values of $m^2$ and $\lambda$. We fix $\alpha_B=1$, $\alpha_H=1.001$ and $\beta_K=3.9993$. The correct amplitude of the scalar power spectrum is fixed by $h_{\mathrm{ds}}=   10^{-5}$ and $f_2=8.8$ while for the perturbations we take $\lambda= 5 \times 10^{-43}$, $m^2= -1.5 \times 10^{-26}$, corresponding to the cosine potential parameters $f/M=1.73 \times 10^8$ and $\mu/\Lambda=0.0046$. We obtain a DHOST tensor-to-scalar ratio of $r_{\mathrm{DHOST}}=10^{-3}$ and $\bar{c}_s=0.976$, while in the perturbed model we find $A_s= 2.76 \times 10^{-9}$, $n_s=0.96716$, $\alpha_s=0.00065$, $\beta_s=-0.000022$ and $r=3.9 \times 10^{-4}$ at the pivot scale. The tensor spectral index has a qualitatively different behaviour with respect to the model presented in Fig. \ref{fig:allPT}, while the consistency relation is closer to the slow roll case. The full results are presented in Fig. \ref{fig:allPTsmallr} and the potential considered together with plots describing the smallness of the perturbations with respect to the background and the condition $\varphi \ll f$, from Eq. (\ref{crucial}) are represented in Fig. \ref{fig:rollpotr}.
As for the previous model, we use Eqs. (\ref{alpha-i}) and (\ref{beta-K}) to determine the derivatives of the functions $f_i$ at $\rm{x}=-1$,

\begin{align}
&f_{2,\rm{x}}=8.809, \qquad f_{1,\rm{x}} = -1.76 \times 10^{-7}, \qquad f_{0,\rm{x}} = -1.05 \times 10^{-9} \nonumber \\ 
&f_{2,\rm{xx}}= 8.8 \beta_H, \qquad f_{1, \rm{xx}} = -0.000088 \beta_B, \qquad f_{0, \rm{xx}}= 2.64\times 10^{-9} \left(\beta_B-4 \beta_H-4.0033\right)  \, .
\end{align}
As before, the range of the coefficients is quite wide. More investigations of the parameter space should be devoted to studying the naturalness of these choices. This is left for future work. 

\subsection{Non-Gaussianities}

We will estimate the level of non-Gaussianities in DHOST models perturbed by a potential term. For this we have to expand the action in terms of the curvature perturbation to cubic order \cite{Chen:2010xka}
\begin{align}
   S_3= \int d\eta &(\prod_{i=1}^3 \slashed{d}^3\tilde k_i) \slashed{\delta} (\vec {\tilde k}_1 + \vec {\tilde k}_2 + \vec {\tilde k}_3) a^2 (C_0\zeta (\tilde k_1) \zeta (\tilde k_2) \zeta (\tilde k_3) + C_1\zeta^\prime (\tilde k_1) \zeta (\tilde k_2) \zeta (\tilde k_3) \nonumber \\ 
   &+C_2\zeta^\prime (\tilde k_1) \zeta^\prime (\tilde k_2) \zeta (\tilde k_3)
   +C_3\zeta^\prime (\tilde k_1) \zeta^\prime (\tilde k_2) \zeta^\prime (\tilde k_3))
\end{align}
where we have expanded the action in Fourier space. The coefficients $C_i$ are {typically} functions of the momenta $\tilde k_i $, {but in the cases such as the ones investigated here, where the potential exhibits an explicit time dependence, they can also depend on time}. There are only {four} possible operators depending on the number of time derivatives acting on $\zeta$. Any second derivative can be brought back to first order by integration by parts. The factor $a^2$ is conventional and comes from the volume of integration.
Typically we are interested in the three point functions
\begin{equation}
    \langle 0\vert \zeta (\tilde k_1) \zeta(\tilde k_2) \zeta (\tilde k_3)\vert 0\rangle = - i\int d\eta\langle 0\vert  [\zeta (\tilde k_1) \zeta(\tilde k_2) \zeta (\tilde k_3), H_3 ]\vert 0\rangle \, ,
\end{equation}
where the interaction picture Hamitonian is given by \cite{Weinberg:2005vy}
\begin{align}
H_3=-\int (\prod_{i=1}^3 \slashed{d}^3\tilde k_i) & \slashed{\delta} (\vec {\tilde k}_1 + \vec {\tilde k}_2+ \vec {\tilde k}_3) a^2 (C_0\zeta (\tilde k_1) \zeta (\tilde k_2) \zeta (\tilde k_3)  + C_1\zeta^\prime (\tilde k_1) \zeta (\tilde k_2) \zeta (\tilde k_3) \nonumber \\
&+ C_2\zeta^\prime (\tilde k_1) \zeta^\prime (\tilde k_2) \zeta (\tilde k_3)
+ C_3\zeta^\prime (\tilde k_1) \zeta^\prime (\tilde k_2) \zeta^\prime (\tilde k_3)) \,.
\end{align}
Using the two-point function $\langle 0\vert \zeta(\tilde k_1) \zeta (\tilde k_2) \vert 0\rangle\simeq  \frac{P_\zeta}{\tilde k_1^3} \slashed{\delta}(\vec{\tilde k}_1+ \vec{\tilde k}_2)$ and Wick's theorem we get the estimate 
\begin{equation}
    \langle 0\vert \zeta (\tilde k_1) \zeta(\tilde k_2) \zeta (\tilde k_3)\vert 0\rangle\simeq \sum_{i=1}^4\langle 0\vert \zeta (\tilde k_1) \zeta(\tilde k_2) \zeta (\tilde k_3)\vert 0\rangle_i \, ,
 \end{equation}
where the contribution to the non-Gaussianities from each of the four operators is 
\begin{equation}
  \langle 0\vert \zeta (\tilde k_1) \zeta(\tilde k_2) \zeta (\tilde k_3)\vert 0\rangle_i\simeq   
    \frac{a}{h_{\rm dS}}C_i  \frac{P_\zeta^3}{\tilde k_1^3 \tilde k_2^3 \tilde k_3^3} \slashed{\delta} (\vec {\tilde k}_1 + \vec {\tilde k}_2+ \vec {\tilde k}_3) \,.
\end{equation}
We have assumed that the coefficients $C_i$ are nearly constant in time. 
After integration over the momenta we get for the cubic moment upon  using $\zeta'\simeq \zeta$,
\begin{equation}
    \langle 0\vert \zeta^3\vert 0\rangle_i\simeq C_i\frac{c_s}{k^2 h_{\rm dS}^2} P_\zeta^3
\end{equation}
from which we can identify $\langle 0\vert \zeta^3\vert 0\rangle_i\simeq 
f_{\rm{ NL}}^i P_\zeta^2$ as
\begin{equation}
    f_{\rm NL}^i\simeq C_i\frac{c_s}{\tilde k^2 h_{\rm dS}^2} P_\zeta \,.
    \label{est1}
\end{equation}
Of course, we have  not determined the exact pre-factors for each of the four operators. Nevertheless, this allows us to extract the order of magnitude of non-Gaussianities \cite{Chen:2010xka}. For example,
in the case of slow-roll inflation where $P_\zeta\simeq \frac{h_{\rm dS}^2}{\epsilon}$
and $\epsilon$ is the first slow-roll parameter, $\epsilon=-\frac{\dot H}{H^2}$, the leading contribution comes from $ \zeta' (\partial \zeta)^2$ with $C_1\simeq  \tilde k^2 \epsilon^2$.  We retrieve then $f_{\rm NL}^1 \simeq \epsilon$. For $P(X,\phi)$ inflation the main contribution comes from the same operator with 
$P_\zeta\simeq \frac{h_{\rm dS}^2}{c_s\epsilon}$ and $C_1\simeq \epsilon k^2 (c_s^{-2}-1)$, we retrieve $f_{\rm NL}^1\simeq (c_s^{-2}-1)$. Similar estimates can be used for other operators \cite{Pirtskhalava:2015zwa}. 

\begin{figure}[!h]
\centering
\includegraphics[width=0.8 \linewidth]{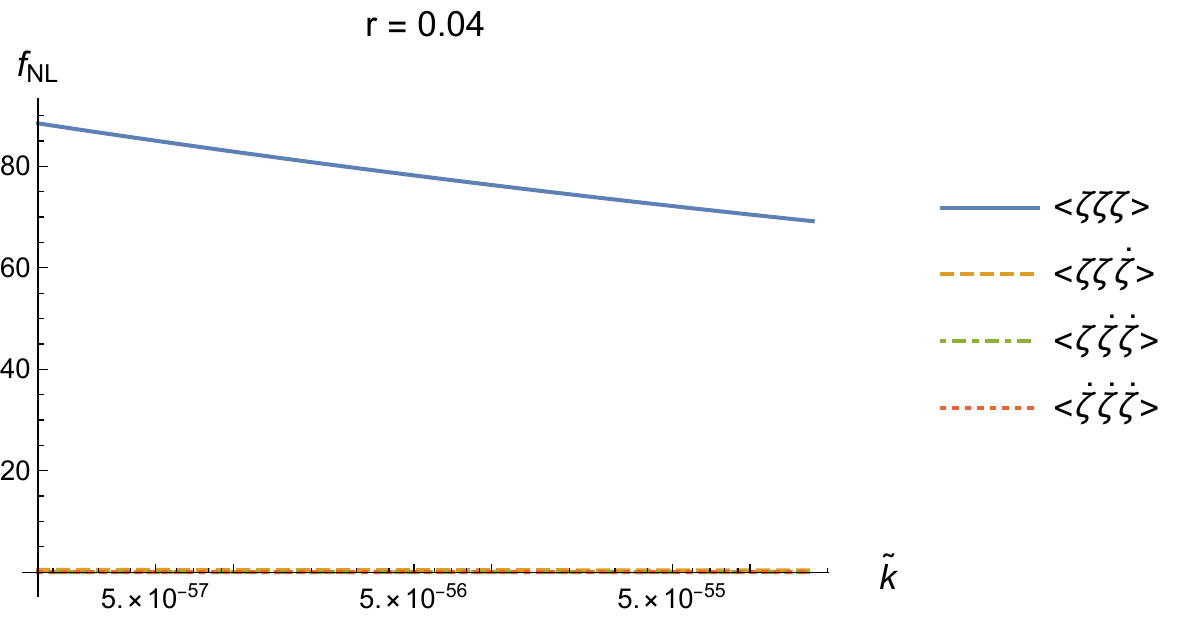} \\
\vspace{8.00mm}
\includegraphics[width=0.8 \linewidth]{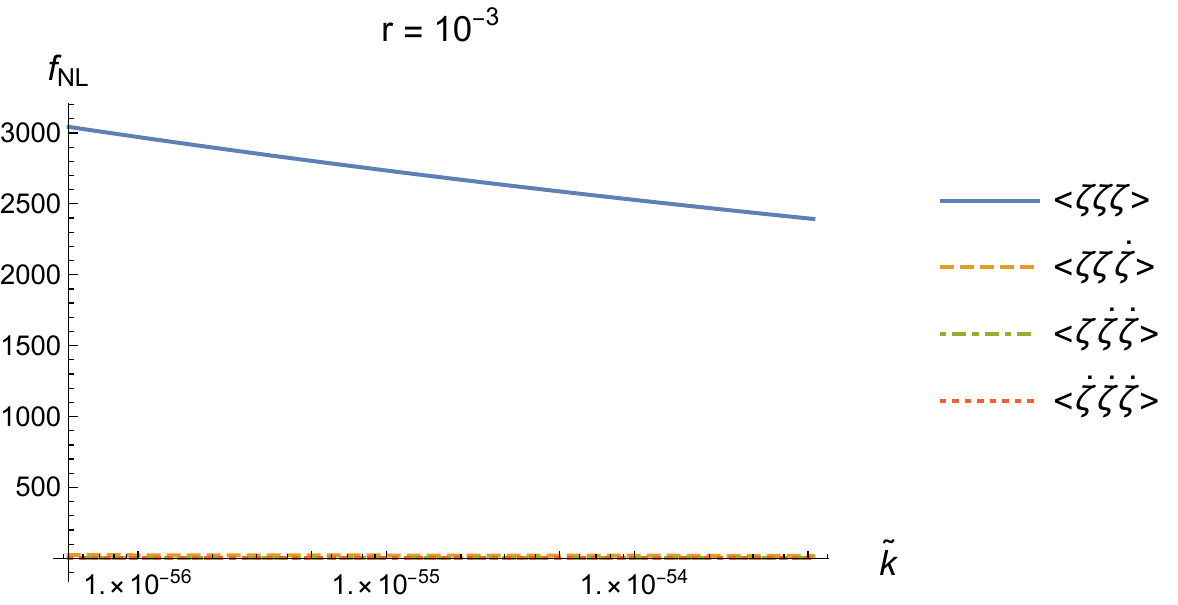} 
\caption{Estimates of the amplitude of the non-Gaussianities generated for the two models with different values of $r$ and  for each of the four operators. The parameters used are $f_{2,{\rm xxx}}=  0$, $f_{0,{\rm xxx}} = 0$,  $f_{1, {\rm xxx}} = 0.00306061$, $f_1 = -0.00035587$, $\beta_B = 0$, $\beta_H = 3.01142$ and $f_{2,{\rm xxx}}= 0$, $f_{0,{\rm xxx}}= 0$,  $f_{1,{\rm xxx}} = 0.00281867$, $f_1 = -0.000404825$, $\beta_B = 0$, $\beta_H = 3.01545$ respectively. Notice that the tensor to scalar ratio in these models cannot be much smaller than the one compatible with the \textit{Planck} 2018 data as the level of non-Gaussianities would exceed the present bound. 
}
\label{fig:fnl}
\end{figure}

When the coefficients $C_i$ depend on time, the integration over $\eta$ can still be done analytically, but the explicit expressions  differ from (\ref{est1}). We find now
\begin{equation}
    f_{\rm NL}^i\simeq \int_{-\infty}^{-\frac{1}{k c_s}} d\eta a^2 C_i(\eta,\tilde k)\frac{ P_\zeta}{\tilde k^3}.
\end{equation}
where the coefficients $C_i$ depend on the conformal time and are evaluated for nearly equilateral configurations $\tilde k\simeq \tilde k_1 \simeq \tilde k_2\simeq \tilde k_3$, see the appendix \ref{fnl:coef}.
We have plotted the estimates for $f_{\rm NL}$ corresponding to the four operators for the two models with a scalar to tensor ratio $r$ compatible with \textit{Planck} and a lower one around $10^{-3}$. {The magnitudes of the predicted non-Gaussianities depend on the values of  six free parameters $\beta_{B},\beta_H, f_1, f_{0,{\rm xxx}}, f_{1,{\rm xxx}}$ and $f_{2,{\rm xxx}}$. There are enough degrees of freedom to set the coefficients $C_2$ and $C_3$ to 0, while the other two yield non-zero contributions. For the first model we have obtained $f_{\rm NL}^{\rm equilateral} \sim  \mathcal{O}(80)$ (Fig. \ref{fig:fnl}, top panel), still within the \textit{Planck} bounds \cite{Akrami:2019izv}, while for the second model we find that significantly larger non-Gaussianities are generated, up to $f_{\rm NL}^{\rm equilateral} \sim  \mathcal{O}(3000)$ (Fig. \ref{fig:fnl}, bottom panel)}. The significant increase of the amplitude of the non-Gaussianity can be understood by looking at the form of $r$ (Eq. \ref{r:DHOST}), in the absence of perturbations. By expressing $\beta_K$ in terms of $\alpha_B$ and $r$ and using the approximation $\bar{c}_s \approx 1$, we notice that $f_{\rm NL}^{\rm equilateral} \sim r^{-3/2}$. As a result, in these models the scalar to tensor ratio $r$ cannot be much smaller than the \textit{Planck} bound as this would lead to larger values for the non-Gaussianities than the ones bound by \textit{Planck} 2018.  As a result the precise calculation of non-Gaussianities in these models is extremely relevant. We leave a detailed computation of the non-Gaussianities generated by each of the operators appearing in such models to a forthcoming paper; in Appendix \ref{fnl:coef} we give  the coefficients $C_0-C_3$ used in our computation.

\subsection{The field excursion and the trans-Planckian censorship conjecture}

\begin{figure}[!h]
\centering
\includegraphics[width=0.49\linewidth]{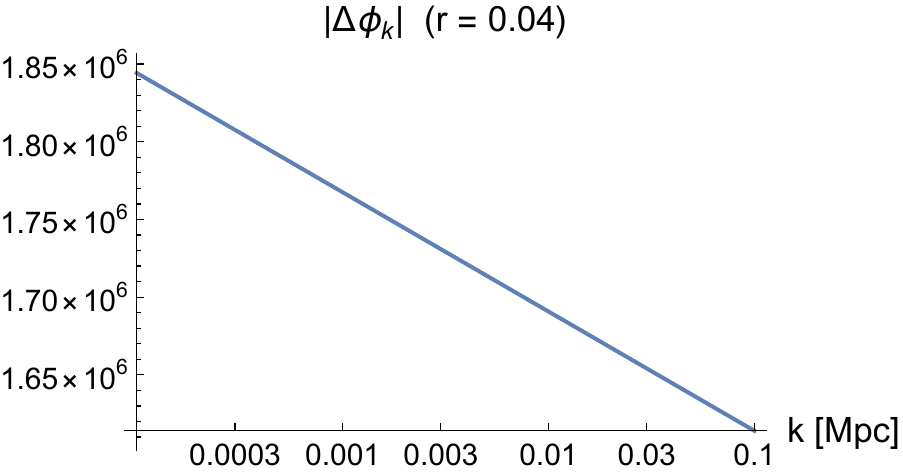} 
\includegraphics[width=0.49\linewidth]{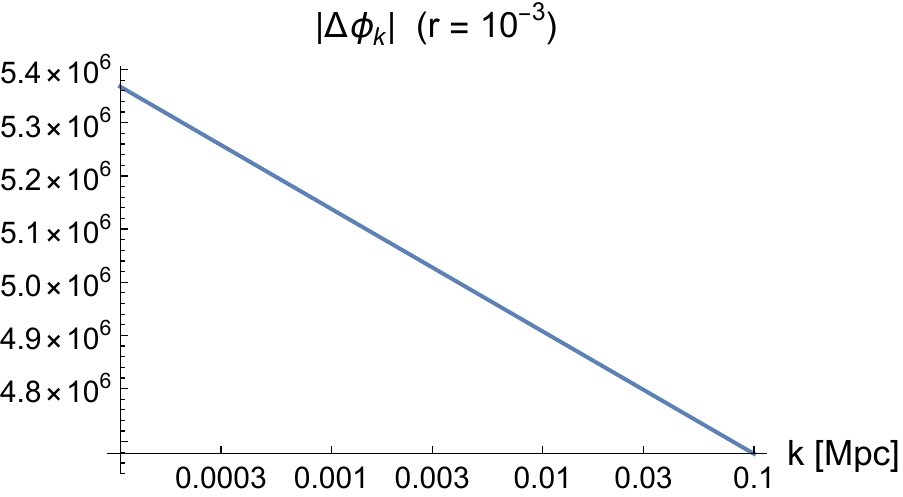} 
\caption{The excursion $\Delta \phi_k$ corresponding to the first model considered as a function of scale, for the observable $k$ range (left) and for the second model (right). Notice that the excursion is shorter than the scale $f$ determining the field scale of the axion-like potential.}
\label{fig:deltaphi}
\end{figure}

In our model inflation is generated by the motion of the scalar field with a constant velocity. Physically this cannot last forever and inflation must end. This could happen for instance if the field contents of the model is more complex and the scalar field $\phi$ triggers an abrupt change of physics when passing a threshold like in the hybrid inflation scenario. We will not study the details of the end of inflation which are left for future work. On the other hand, one can garner information using simple estimates. First of all as an order of magnitude the scale factor at the end of inflation is
\begin{equation}
    a_{\rm end}\simeq \left(\frac{H_0}{H_{end}}\right)^{1/2} =\left(\frac{H_0}{h_{\rm dS}  m_{\rm Pl}}\right)^{1/2}
\end{equation}
where we have normalised the scale factor in the present Universe as $a_0=1$ and we have neglected the matter era duration compared to the radiation one.
Notice that $H_{\rm end}\simeq H$ as we take inflation to be in a near de Sitter phase here. 
This leads for the number of efoldings between the time when the pivot scale $k_\star$ enters the horizon and the end of inflation
\begin{equation}
N_\star= \ln \left(\frac{a_{\rm end}H_{\rm end}}{k_\star}\right).
\end{equation}
For the models that we have considered we find $N_\star= 59.52 $ for the first model with $r\sim 0.04$ and $N_\star=58.97$ for the second one with $r\simeq 10^{-3}$.
An important quantity is the excursion
\begin{equation}
 \Delta \phi_k\equiv \vert \phi(t_{\rm end}) - \phi(t_k) \vert
\end{equation}
corresponding to the change of the scalar field between horizon entry and the end of inflation. We have represented in Fig. \ref{fig:deltaphi} the excursion $\Delta \phi_k$ as a function of scale for the two models considered. As we can see the excursion is much smaller  than the Planck scale for all the observable scales as long as $ M\lesssim 10^{-6} m_{\rm Pl}$. As $M$ is the only free parameter in our model which is not constrained by the CMB data, we can impose this bound as a theoretical requirement \cite{Vafa:2005ui, Bedroya:2019snp}. This implies that  the distance conjecture of string theory is fulfilled \cite{Ooguri:2006in}. In particular we can see in Fig. \ref{fig:deltaphi} that the values of the potential term at horizon exit for the physical scales is always much smaller than the energy density of inflation. As the energy scale of inflation is much smaller than $m_{\rm Pl}^4$, quantum gravity effects in these models are negligible.

Another issue is the Trans-Planckian nature of some of the modes in slow roll inflation \cite{Brandenberger:2012aj}. This does not  happen if one requires that all modes of length scale the Planck scale $l_{\rm Pl}= m^{-1}_{\rm Pl}$ and below satisfy
\begin{equation}
    \frac{a(t_{\rm end})}{a_{\rm in}} l_{\rm Pl}< H^{-1}
\end{equation}
where $H$ is constant here. This simply expresses that modes which are sub-Planckian never become super-Hubble and therefore become eventually classicalised. In our case this constraint imposes that the total number of e-folds of inflation must be bounded 
\begin{equation}
   N_T=\ln \frac{a_{\rm end}}{ a_{\rm in}} < -\ln ( h_{\rm dS}) \, .
\end{equation}
We find $N_T$ is less than  $10.41$ and $11.51$ respectively. This is much smaller than $N_\star$.
As $h_{\rm dS}$ is determined by the COBE normalisation, we conclude that  the DHOST model do not  evade the trans-Planckian issue.  The absence of amplification for the trans-Planckian modes would  require a modification of our scenario which is beyond the present work.
\section{Conclusions}

In this paper we show that small perturbations around DHOST de Sitter space-times  are able to describe inflationary models compatible with the latest \textit{Planck} constraints. In these models, the usual paradigm of inflation driven by the inflaton's scalar potential is modified. The  interaction potentials only serve to generate the appropriate power spectra of perturbations whilst the de Sitter nature of space-time results from the DHOST action.  In particular, we find that simple  perturbations of the type $m^2 \phi^2$ and $\lambda \phi^4$ yield a nearly-scale invariant power spectrum of curvature perturbations, and that the free parameters $\lambda$ and $m^2$ can be tuned such that the model is compatible with the inflationary constraints on $n_s$, $\alpha_s$ and $\beta_s$. Parameters from these class of models can also be fixed to predict a tensor-to-scalar ratio compatible with the latest constraints, and also  future ones. 

So far we have not mentioned the necessary end of inflation. As such this phenomenon is not included in the treatment of inflation that we have given. One possibility for the end of inflation would be that the polynomial interactions which are subdominant in our approach could become eventually dominant and lead to a violation of the usual slow-roll conditions. This would eventually lead to the end of inflation. Another possibility could be the presence of a transition to another regime of the model, as happens in hybrid inflation \cite{Linde:1993cn}, where beyond a threshold the scalar could trigger the opening of a new valley in field space.  We leave these issues open for the future. Another topic that we have not covered is the radiative stability of the models. As we have seen, some of the parameters of the models need to be adjusted, maybe even finely adjusted (see $\alpha_H$ for instance). Quantum corrections in these model may lead to a detuning of these parameters although it is expected that DHOST theories obey a non-renormalisation theorem akin to the one for $K(X)$-models which would render quantum issues easier to handle \cite{deRham:2014wfa,Brax:2016jjt}. Finally, non-Gaussianities could turn out to be crucial in the inflationary models presented here in comparison with data as they could be substantial. We have given estimates for the level of non-Gaussianities generated by the cubic operators in the curvature perturbation. We have found that they can be made compatible with the current \textit{Planck} 2018 bounds provided the tensor to scalar ratio is also compatible with the same data but not extremely small. Indeed when the ratio $r$ is much smaller, we find that the level of non-Gaussianities could exceed the present bounds. This implies that the models considered in this paper are within reach of the next generation of CMB experiments. Of course, a more detailed analysis of the non-Gaussianities and its dependence on the parameter of the models must be performed to confirm these estimates. This is left for future work. 

\appendix
\section{Other perturbation operators}
\label{appendixLi}
In the previous sections, we have considered perturbations to the DHOST actions of the $L_2$ type (Eq. \ref{eq:scodL2}). The other types of scordatura corrections that can be considered are those proportional to the $L_1$-$L_5$ Lagrangians, and they will produce second-order perturbations that can be cast in a similar fashion to Eq. (\ref{Lagrangian-scordatura}) of section \ref{sec:scod_pert}, but with different coefficients. These coefficients are presented in the next few paragraphs. As we have noted in Section \ref{sec:scod_pert}, in the case of the $L_4$ correction, we obtain an additional term and in this case the second order action (\ref{Lagrangian-scordatura-red}) gets an additional term,
\begin{equation}\label{Lagrangian-scordatura-appendix}
{\tilde {\cal L}}_{\rm S}^{(2)} = \frac{a^3}{2} \bigg[ \Big( {\cal A}_1 
+ \frac{a^2 {\cal A}_{2}}{{\tilde k}^2+\alpha {k}_{\rm IR}^2}
+ \frac{{\tilde k}^2 {\cal A}_{3}}{a^2} \Big) \, \dot{\zeta}^2 
- \Big( {\cal B}_{1} \Big(\frac{{\tilde k}}{a}\Big)^2 + {\cal B}_{2}
\Big(\frac{{\tilde k}}{a}\Big)^4 + {\cal M}\Big) \,\zeta^2
\bigg] \,,
\end{equation}
leading to a modification of ${\cal K}$ and $c_{\rm s}^2$ to
\begin{align}
{\cal K} &\equiv \bar{\cal A} 
\bigg( 1 + \frac{\alpha}{2f_2} \Big( \frac{{\cal A}_{1}}{\bar{\cal A}} 
+ \frac{a^2}{{\tilde k}^2+\alpha k_{\rm IR}^2} \frac{{\cal A}_{2}}{\bar{\cal A}} +\frac{{\tilde k}^2 }{a^2} \frac{{\cal A}_{3}}{\bar{\cal A}}\Big)
\bigg) \,, \\
c_{\rm s}^2 ({\tilde k}) &\equiv \bar{c}_{\rm s}^2 
+ \frac{\alpha]}{2f_2} \bigg(
\frac{{\cal B}_{1}}{\bar{\cal A}} - \bar{c}_{\rm s}^2 \frac{{\cal A}_{1}}{{\cal A}}
+ \Big(\frac{{\tilde k}}{a}\Big)^{2} \frac{{\cal B}_{2}}{\bar{{\cal A}}}
- \bar{c}_{\rm s}^2 \Big(\frac{{\tilde k}}{a}\Big)^{2} \frac{{\cal A}_{3}}{\bar{{\cal A}}}
\bigg) \,.
\end{align}
\subsection{$L_1$}
Here 
\begin{equation}
\label{eq:scodL1}
S_{\rm{S}_1} = \int d^4 x \sqrt{-g} \bigg[ - \frac{\alpha}{2} \frac{\phi_{\nu\eta} \phi^{\nu\eta}}{M^2} \bigg] \,
\end{equation}
and

\begin{align}
& {\bar k}_{11} = 3 a^3 {\rm x} , \hspace{2.5cm} 
{\bar k}_{12} = - 3 a^3 {\rm x} \alpha_H  , \hspace{2.5cm} 
{\bar k}_{22} = a^3 {\rm x} (1+3\alpha_H^2)  , \\ \nonumber 
& {\bar n}_{12} = 3 a^3 \dot{\varphi} \big( 3 h_b \dot{\varphi} 
+ (-1+5\alpha_H + 2 \alpha_{H}^2-2\beta_H)  \ddot{\varphi} \big) , \hspace{.5cm} 
{\bar n}_{13} =  a {\rm x} , \hspace{.5cm} 
{\bar n}_{23} = - a {\rm x} \alpha_H , \\ \nonumber 
& {\bar m}_{11}  =  \frac{3}{2} a^3 \Big( {\rm x} ( 2\dot{h}_b  + 3 h_b^2 ) 
- 2 ( 2 + 3 \alpha_H ) h_b \dot{\varphi} \ddot{\varphi}
+ ( 1 - 6 \alpha_H - 7 \alpha_H^2 
+ 4 \beta_H ) \ddot{\varphi}^2 
- 2  \alpha_H  \dot{\varphi} \dddot{\varphi} \Big) , 
\\ \nonumber 
& {\bar m}_{12} =  \frac{3}{2} a^3 \Big( \left(7 \alpha_H^3-3 \alpha_H^2+\alpha_H (5-4 \beta_H)-1\right) \ddot{\varphi}^2+\dot{\varphi}^2 \left(2 \alpha_H \dot{h}_b+3 (\alpha_H-3) h_B^2\right)  \nonumber \\
& \hspace{2cm} +2 \dot{\varphi} \left(\left(\alpha_H^2+1\right) \dddot{\varphi} +\left(3 \alpha_H^2-7 \alpha_H+3\right) h_b \ddot{\varphi}\right)
\Big) ,  \nonumber \\ 
\nonumber
& {\bar m}_{22} = 
- \frac{1}{2} a \Big( 
-24 a^2 \alpha_H f^{(3)} x^3 \left(\ddot{\varphi}\right)^2-120 a^2 \alpha_H^2 \beta_H f \left(\ddot{\varphi}\right)^2+18 a^2 \alpha_H^3 f \dddot{\varphi} \dot{\varphi}+30 a^2 \alpha_H^2 f \dddot{\varphi} \dot{\varphi} \\ \nonumber
& \hspace{2cm}
+105 a^2 \alpha_H^4 f \left(\ddot{\varphi}\right)^2+138 a^2 \alpha_H^3 f \left(\ddot{\varphi}\right)^2+36 a^2 \alpha_H^2 f \left(\ddot{\varphi}\right)^2-12 a^2 \alpha_H \beta_H f \dddot{\varphi} \dot{\varphi} \\ \nonumber
& \hspace{2cm}
-84 a^2 \alpha_H \beta_H f \left(\ddot{\varphi}\right)^2+6 a^2 \alpha_H f \dddot{\varphi} \dot{\varphi}+30 a^2 \alpha_H f \left(\ddot{\varphi}\right)^2-12 a^2 \beta_H f \left(\ddot{\varphi}\right)^2 \nonumber \\
& \hspace{2cm}
-6 a^2 \alpha_H^2 f x h_b'-9 a^2 \alpha_H^2 f h_b^2 \mathrm{x}-18 a^2 \alpha_H f \mathrm{x} h_b'+45 a^2 f h_b^2 \mathrm{x}+54 a^2 \alpha_H^3 f h_b \dot{\varphi} \ddot{\varphi} \nonumber \\
& \hspace{2cm}
+12 a^2 \alpha_H^2 f h_b \dot{\varphi} \ddot{\varphi}-36 a^2 \alpha_H \beta_H f h_b \dot{\varphi} \ddot{\varphi}-72 a^2 \alpha_H f h_b \dot{\varphi} \ddot{\varphi}+36 a^2 \beta_H f h_b \dot{\varphi} \ddot{\varphi} \nonumber  \\
& \hspace{2cm}
+30 a^2 f h_b \dot{\varphi} \ddot{\varphi}+10 a^2 f \dddot{\varphi} \dot{\varphi}-5 a^2 f \left(\ddot{\varphi}\right)^2-4 f k^2 \rm{x}
\Big) ,
 \nonumber \\
& {\bar m}_{22{\rm s}}=0 \,, \hspace{0.7cm} {\bar m}_{23} = -  a \Big( - 3 h_b {\rm x} 
+ \big( -1 + 5 \alpha_H + 2 \alpha_H^2 - 2 \beta_H \big) \dot{\varphi} \ddot{\varphi}  \Big) \,, \hspace{0.7cm} 
{\bar m}_{33{\rm s}} = \frac{{\rm x}}{a} \,, \nonumber \\ \nonumber 
& {\bar m}_{33} = \frac{1}{2} a \Big( - 9 h_b^2 {\rm x}
+ \big( 1 - 3 \alpha_H \big)^2 \ddot{\varphi}^2
- 2 \dot{\varphi} \big( 3  h_b (1-3 \alpha_H)\ddot{\varphi}+ \dddot{\varphi} \big) \Big) \,.
\end{align}

\subsection{$L_3$}
\begin{equation}
\label{eq:scodL3}
S_{\rm{S}_3} = \int d^4 x \sqrt{-g} \bigg[ - \frac{\alpha}{2} \frac{\Box\phi \, \phi_{\nu}\phi^{\nu\eta} \phi_{\eta}}{M^2} \bigg] \,
\end{equation}

and

\begin{align}
& {\bar k}_{11} = 0 , \hspace{2.5cm} 
{\bar k}_{12} = - \frac{3}{2} a^3 {\rm x}^2  , \hspace{2.5cm} 
{\bar k}_{22} = a^3 {\rm x}^2 (1+3\alpha_H) , \\ \nonumber
& {\bar n}_{12} =  \frac{9}{2} a^3 \big( {\rm x}^2 h_b 
+ (1-\alpha_H){\rm x} \dot{\varphi} \ddot{\varphi} \big) , \hspace{.5cm}
{\bar n}_{13} = 0 , \hspace{.5cm} 
{\bar n}_{23} = - \frac{1}{2} a {\rm x}^2 , \\ \nonumber
& {\bar m}_{11}  = - \frac{3}{2} a^3 {\rm x} \Big( 2 \ddot{\varphi}^2 
+  \dot{\varphi} \dddot{\varphi} \Big) , 
\\ \nonumber 
& {\bar m}_{12} = \frac{3}{2} a^3 {\rm x} \left(-3 {\rm x} \left(\dot{h}_b+3 h_b^2\right)+\dot{\varphi} \left(3 (6 \alpha_H h_b+h_b) \ddot{\varphi}+2 (2 \alpha_H+1) \dddot{\varphi}\right)+\left(15 \alpha_H^2+8 \alpha_H-6 \beta_H+1\right) \ddot{\varphi}^2\right) , \\ \nonumber
& {\bar m}_{22} = 
- \frac{1}{2f} a^3 {\rm x} \Big( \ddot{\varphi}^2 \left(12 f^{(3)} {\rm x}^3-f \left(105 \alpha_H^3+201 \alpha_H^2+\alpha_H (87-90 \beta_H)-78 \beta_H+7\right)\right) \\ \nonumber 
& \hspace{2cm} 
+3 (3 \alpha_H+7) f {\rm x} \left(\dot{h}_b+3 h_b^2\right)-f \dot{\varphi} \left(3 h_b \left(30 \alpha_H^2+57 \alpha_H-12 \beta_H+7\right) \ddot{\varphi} \right. \\ \nonumber
& \hspace{2cm} 
+\left. 2 \dddot{\varphi} \left(9 \alpha_H^2+24 \alpha_H-3 \beta_H+7\right)\right) \Big) ,
\\ \nonumber
& {\bar m}_{22{\rm s}}=0 \,, \hspace{0.7cm} {\bar m}_{23} = - \frac{3}{2}  a  \Big(  h_b {\rm x}^2
+ \big( 1 -  \alpha_H) {\rm x}  \dot{\varphi} \ddot{\varphi}  \Big) \,, \hspace{0.7cm}
{\bar m}_{33{\rm s}} = 0 \,, \nonumber \\ \nonumber
& {\bar m}_{33} = -\frac{1}{2} a {\rm x} \left(-3 {\rm x} \left(\dot{h}_b+3 h_b^2\right)+\dot{\varphi} \left(3 (6 \alpha_H h_b+h_b) \ddot{\varphi}+(3 \alpha_H+2) \dddot{\varphi}\right)+\left(15 \alpha_H^2+6 \alpha_H-6 \beta_H+1\right) \ddot{\varphi}^2\right) \,.
\end{align}

\subsection{$L_4$}
\begin{equation}
\label{eq:scodL4}
S_{\rm{S}_4} = \int d^4 x \sqrt{-g} \bigg[ - \frac{\alpha}{2} \frac{\phi^{\nu} \phi_{\nu\eta} \phi^{\eta\lambda}\phi_{\lambda}}{M^2} \bigg] \,
\end{equation}

\begin{align}
& {\bar k}_{11} = 0 , \hspace{2.5cm} 
{\bar k}_{12} = 0  , \hspace{2.5cm} 
{\bar k}_{22} = a^3 {\rm x}^2  , \\ \nonumber
& {\bar n}_{12} = - 3 a^3 {\rm x} \dot{\varphi} \ddot{\varphi} , \hspace{2.5cm}
{\bar n}_{13} = 0 , \hspace{2.5cm} 
{\bar n}_{23} = 0 , \\ \nonumber
& {\bar m}_{11}  =  \frac{3}{2} a^3 {\rm x}  \ddot{\varphi}^2  , 
\\ \nonumber 
& {\bar m}_{12} = \frac{3}{2} a^3 {\rm x} \left( (1+5\alpha_H) \ddot{\varphi}^2+ 2 \dot{\varphi}(3 h_b \ddot{\varphi}+\dddot{\varphi}) \right) , \\ \nonumber
& {\bar m}_{22} = 
- \frac{1}{2} a^3 {\rm x} \Big( \left(7+42 \alpha_H+ 27 \alpha_H^2-12 \beta_H \right) \ddot{\varphi}^2 + 2 \dot{\varphi} (7+3\alpha_H)\left(3 h_b \ddot{\varphi} +\dddot{\varphi}\right)  
 \Big) ,
\\ \nonumber
& {\bar m}_{22{\rm s}}=a {\rm x}^2 \,, \hspace{0.7cm} {\bar m}_{23} =   a  {\rm x}  \dot{\varphi} \ddot{\varphi}  \,, \hspace{0.7cm}
{\bar m}_{33{\rm s}} = 0 \,, \nonumber \\ \nonumber
& {\bar m}_{33} = -\frac{1}{2} a {\rm x} \left( (1+6 \alpha_H) \ddot{\varphi}^2+ 2 \dot{\varphi} (3 h_b \ddot{\varphi} + \dddot{\varphi}) \right) \,.
\end{align}

\subsection{$L_5$}
\begin{equation}
\label{eq:scodL5}
S_{\rm{S}_5} = \int d^4 x \sqrt{-g} \bigg[ - \frac{\alpha}{2} \frac{(\phi_{\nu}\phi^{\nu\eta}\phi_{\eta})^2}{M^2} \bigg] \,
\end{equation}

\begin{align}
& {\bar k}_{11} = 0 , \hspace{2.5cm} 
{\bar k}_{12} = 0  , \hspace{2.5cm} 
{\bar k}_{22} = a^3 {\rm x}^3  , \\ \nonumber
& {\bar n}_{12} = - 3 a^3 {\rm x}^2 \dot{\varphi} \ddot{\varphi} , \hspace{2.5cm}
{\bar n}_{13} = 0 , \hspace{2.5cm} 
{\bar n}_{23} = 0 , \\ \nonumber
& {\bar m}_{11}  =  \frac{3}{2} a^3 {\rm x}^2  \ddot{\varphi}^2  , 
\\ \nonumber 
& {\bar m}_{12} = \frac{3}{2} a^3 {\rm x}^2 \left( (3+5\alpha_H) \ddot{\varphi}^2+ 2 \dot{\varphi}(3 h_b \ddot{\varphi}+\dddot{\varphi}) \right) , \\ \nonumber
& {\bar m}_{22} = 
- \frac{3}{2} a^3 {\rm x}^2 \Big( \left(9+18 \alpha_H+ 9 \alpha_H^2-4 \beta_H \right) \ddot{\varphi}^2 + 2 \dot{\varphi} (3+\alpha_H)\left(3 h_b \ddot{\varphi} +\dddot{\varphi}\right)  
 \Big) ,
\\ \nonumber
& {\bar m}_{22{\rm s}} = 0 \,, \hspace{0.7cm}
{\bar m}_{23} =   a  {\rm x}^2  \dot{\varphi} \ddot{\varphi}  \,, \hspace{0.7cm}
{\bar m}_{33{\rm s}} = 0 \,, \nonumber \\ \nonumber
& {\bar m}_{33} = -\frac{1}{2} a {\rm x}^2 \left( 3(1+2 \alpha_H) \ddot{\varphi}^2+ 2 \dot{\varphi} (3 h_b \ddot{\varphi} + \dddot{\varphi}) \right) \,.
\end{align}

\section{Coefficients of the four operators yielding non-Gaussianities}
\label{fnl:coef}
Here we provide the coefficients of the four cubic operators appearing in our shift-symmetry breaking model. The coefficients given are those obtained by expanding the action to third order, and include the $a^2$ factor.

\begin{align}
a^2& C_0 = \frac{1}{48   (\alpha_B+1)^5 h_{\rm dS}^3 k^2}  (-a^2 (c-t)^2 \left(\lambda (c-t)^2+12 m^2\right) ((\alpha_B+1)^3 (\alpha_H+1)^2 h_{\rm dS} k^4 \nonumber \\
&-a^2 (\alpha_B+1)^5 h_{\rm dS}^3 k^2)-24 (\alpha_B+1)^2 k^4 ((\alpha_B+1) (\alpha_H+1) h_{\rm dS} (a^2 (\alpha_B+1) h_{\rm dS} (f_1+6 f_2 h_{\rm dS}) \nonumber \\
&-\frac{32 (\alpha_B+1) \alpha_H f_2 k^2 (\alpha_B-\alpha_H)}{3 \left(\alpha_B^2+2 \alpha_B-\beta_K+1\right)})+2 a^2 (\alpha_B+1)^2 f_2 h_{\rm dS}^3 (\alpha_B-2 \alpha_H-1) \nonumber \\
&+(\alpha_H+1)^3 k^2 (f_1+2 f_2 h_{\rm dS} (2 \alpha_B+4 \alpha_H+3))-2 (3-2 \alpha_H) (\alpha_H+1)^2 f_2 h_{\rm dS} k^2 (\alpha_B-2 \alpha_H-1)) \nonumber \\
&+\frac{1}{\alpha_B^2+2 \alpha_B-\beta_K+1} (8 (\alpha_B+1)^3 k^2 (-(\alpha_H+1)^2 k^2 (9 a^2 h_{\rm dS}^2 \left(\alpha_B^2+2 \alpha_B-\beta_K+1\right) (f_1+6 f_2 h_{\rm dS}) \nonumber \\
&+k^2 (\alpha_B-\alpha_H) (2 f_2 h_{\rm dS} (\alpha_B (6 \alpha_H+2)+7 \alpha_H-9)-(\alpha_H+3) f_1)) \nonumber \\
&+6 a^2 (\alpha_H+1) h_{\rm dS}^2 k^2 (-\alpha_B+2 \alpha_H+1) (\alpha_B^2+2 \alpha_B-\beta_K+1) (f_1+6 f_2 h_{\rm dS}) \nonumber \\
&+3 a^2 (\alpha_B+1)^2 h_{\rm dS}^2 (k^2 (\alpha_B-\alpha_H) ((\alpha_H-1) f_1+2 (\alpha_H-3) f_2 h_{\rm dS}) \nonumber \\
&-3 a^2 h_{\rm dS}^2 \left(\alpha_B^2+2 \alpha_B-\beta_K+1\right) (f_1+2 f_2 h_{\rm dS})))) \nonumber \\
&+96 (\alpha_B+1)^3 f_2 h_{\rm dS} k^4 \left((\alpha_H+1)^2 (2 \alpha_H+3) k^2-6 a^2 (\alpha_B+1)^2 h_{\rm dS}^2\right))
\end{align}

\begin{align}
a^2& C_1 = \frac{1}{48  (\alpha_B+1)^4  h_{\rm dS}^2 k^3 a} (a^2 (\lambda (c-t)^2+12 m^2) (\alpha_B+1) (\alpha_H+1) (k^2 (\alpha_H+1)^2-3 a^2 h_{\rm dS}^2 \nonumber \\
&\times (3 \alpha_B^2+6 \alpha_B-2 \beta_K+3)) (c-t)^2+24 (4 a^2 f_2 k^2 (\alpha_B+1)^2 \alpha_H (\alpha_B-2 \alpha_H-1) h_{\rm dS}^2 \nonumber \\
&+2 a^2 f_2 (\alpha_B+1)^2 (6 a^2 h_{\rm dS}^2 (\alpha_B^2+2 \alpha_B-\beta_K+1)-k^2 (\alpha_B+1) (\alpha_H+1)) h_{\rm dS}^2+12 a^2 f_2 k^2 (\alpha_H+1) \nonumber \\
&\times  (3-2 \alpha_H) (\alpha_B-2 \alpha_H-1) ((\alpha_B+1)^2-\beta_K) h_{\rm dS}^2-3 a^2 (\alpha_B+1) (a^2 h_{\rm dS} (f_1+6 f_2 h_{\rm dS}) (\alpha_B+1) \nonumber \\
&-\frac{32 f_2 k^2 (\alpha_B+1) (\alpha_B-\alpha_H) \alpha_H}{3 (\alpha_B^2+2 \alpha_B-\beta_K+1)}) ((\alpha_B+1)^2-\beta_K) h_{\rm dS}^2-9 a^2 k^2 (\alpha_H+1)^2 (f_1+2 f_2 h_{\rm dS} \nonumber \\
& \times (2 \alpha_B+4 \alpha_H+3)) ((\alpha_B+1)^2-\beta_K) h_{\rm dS}-2 f_2 k^2 (3-2 \alpha_H) (\alpha_H+1)^2 (6 a^2 h_{\rm dS}^2 (\alpha_B^2+2 \alpha_B \nonumber \\
&-\beta_K+1)-k^2 (\alpha_B+1) (\alpha_H+1))-\frac{1}{3 (\alpha_B^2+2 \alpha_B-\beta_K+1)}(2 k^2 (\alpha_B+1)^2 (\alpha_H+1) \nonumber \\
&\times (3 h_{\rm dS} (-\alpha_H f_1+f_1+2 f_2 h_{\rm dS} (8 \alpha_B+21 \alpha_H+11)) (\alpha_B^2+2 \alpha_B-\beta_K+1) a^2+2 f_2 k^2 \nonumber \\
& \times (\alpha_B-\alpha_H) (18 \alpha_H^2-7 \alpha_H+3)))-2 f_2 k^2 (\alpha_B+1) (-36 a^2 (\alpha_B+1)^2 \alpha_H h_{\rm dS}^2-a^2 (\alpha_B+1)^2 \nonumber \\
&\times (4 \alpha_H+3) h_{\rm dS}^2-12 a^2 (\alpha_H+1) (2 \alpha_H+3) ((\alpha_B+1)^2-\beta_K) h_{\rm dS}^2-k^2 (\alpha_H+1)^2 (8 \alpha_H^2-4 \alpha_H+5)) \nonumber \\
& -\frac{1}{k^2 (\alpha_B^2+2 \alpha_B-\beta_K+1)}((\alpha_B+1) (3 h_{\rm dS}^3 k^2 (\alpha_B^2+2 \alpha_B-\beta_K+1) (f_1 (3 \alpha_B^3+(7-2 \alpha_H) \alpha_B^2 \nonumber \\
& + (-4 \alpha_H-2 \beta_K+5) \alpha_B+2 \alpha_H (\beta_K-1)+1)+2 f_2 h_{\rm dS} (7 \alpha_B^3+(15-6 \alpha_H) \alpha_B^2 \nonumber \\
&-3 (4 \alpha_H+2 \beta_K-3) \alpha_B+6 \alpha_H (\beta_K-1)+1)) a^4+h_{\rm dS} k^4 (f_1 (\alpha_H+1) ((\alpha_H-15) \alpha_B^3 \nonumber \\
&+(4 \alpha_H^2+31 \alpha_H-21) \alpha_B^2+(8 \alpha_H^2+(\beta_K+59) \alpha_H+9 \beta_K+3) \alpha_B-9 \beta_K+\alpha_H^2 (4-6 \beta_K) \nonumber \\
&+\alpha_H (29-23 \beta_K)+9)+2 f_2 h_{\rm dS} (2 (9 \alpha_H^2+16 \alpha_H+3) \alpha_B^4-(36 \alpha_H^3+9 \alpha_H^2-26 \alpha_H+49) \alpha_B^3 \nonumber \\
&-(116 \alpha_H^3+(12 \beta_K+79) \alpha_H^2+(8 \beta_K-38) \alpha_H-4 \beta_K+95) \alpha_B^2+(4 (9 \beta_K-31) \alpha_H^3 \nonumber \\
&+(47 \beta_K-59) \alpha_H^2+6 (11 \beta_K+21) \alpha_H+55 \beta_K-19) \alpha_B+\alpha_H^3 (42 \beta_K-44) \nonumber \\
&-3 (2 \beta_K^2+5 \beta_K-7)-2 \alpha_H (6 \beta_K^2+26 \beta_K-41)+\alpha_H^2 (-6 \beta_K^2+5 \beta_K-7))) a^2))))
\end{align}

\begin{align}
a^2& C_2 = \frac{1}{48 h_{\rm dS} k^2 (\alpha_B+1)^5 k^3} (-96 f_2 h_{\rm dS} (\alpha_B+1) (18 (\alpha_B+1)^2 \alpha_H^2 k^2+(\alpha_B+1)^2 \alpha_H \nonumber \\
&\times (4 \alpha_H+3) k^2+(\alpha_B+1)^2 (-3 \alpha_H^2 x+4 \alpha_H-4 \beta_H) k^2+3 (\alpha_H+1) (-8 \alpha_H^2 +4 \alpha_H-5) \nonumber \\
&\times ((\alpha_B+1)^2-\beta_K) k^2-9 a^2 h_{\rm dS}^2 (2 \alpha_H+3) ((\alpha_B+1)^2-\beta_K)^2) k^2+24 (-18 a^2 f_2 (3-2 \alpha_H) \nonumber \\
&\times (-\alpha_B+2 \alpha_H+1) ((\alpha_B+1)^2-\beta_K)^2 h_{\rm dS}^3-27 a^2 (\alpha_H+1) (f_1+2 f_2 h_{\rm dS} (2 \alpha_B+4 \alpha_H+3)) \nonumber \\
& \times ((\alpha_B+1)^2-\beta_K)^2 h_{\rm dS}^2-f_2 (\alpha_B+1)^2 (24 a^2 \alpha_H (\alpha_B^2+2 \alpha_B-\beta_K+1) h_{\rm dS}^2-4 k^2 (\alpha_B+1) \alpha_H \nonumber \\
& \times (\alpha_H+1)+2 k^2 (\alpha_B-2 \alpha_H-1) (\alpha_H^2-12 \alpha_H+4 \beta_H+3)) h_{\rm dS}+12 f_2 (\alpha_H+1) (3-2 \alpha_H) \nonumber \\
&\times (k^2 (\alpha_B+1) (\alpha_H+1)-6 a^2 h_{\rm dS}^2 (\alpha_B^2+2 \alpha_B-\beta_K+1)) ((\alpha_B+1)^2-\beta_K) h_{\rm dS}+(\alpha_B+1)^2 k^2\nonumber \\
&\times ( (\alpha_H+1) (f_1 (-\alpha_H^2+2 \alpha_H+3)+2 f_2 h_{\rm dS} (105 \alpha_H^2+4 \alpha_H+\alpha_B (24 \alpha_H-2)+2 \beta_B-12 \beta_H+17)) \nonumber \\
&-\frac{1}{\alpha_B^2+2 \alpha_B-\beta_K+1}(h_{\rm dS} (6 h_{\rm dS} (f_1+6 f_2 h_{\rm dS}) (\alpha_B^2+2 \alpha_B-\beta_K+1) a^2-3 h_{\rm dS} (f_1 \alpha_H \nonumber \\
&-2 f_2 h_{\rm dS} (16 \alpha_B+45 \alpha_H+16)) (\alpha_B^2+2 \alpha_B-\beta_K+1) a^2+4 f_2 k^2 (\alpha_B-\alpha_H) (10 \alpha_H^2-7 \alpha_H+3) \nonumber \\
&+\alpha_H (32 f_2 k^2 (\alpha_B-\alpha_H) \alpha_H-3 a^2 h_{\rm dS} (f_1+6 f_2 h_{\rm dS}) (\alpha_B^2+2 \alpha_B-\beta_K+1))) ((\alpha_B+1)^2-\beta_K)))) k^2 \nonumber \\
&+3 a^2 h_{\rm dS} (\lambda (c-t)^2+12 m^2) (c-t)^2 (\alpha_B+1) (3 (\alpha_B+1)^2 k^2+3 (\alpha_B+1)^2 \alpha_H^2 k^2+6 (\alpha_B+1)^2 \alpha_H k^2 \nonumber \\
&-2 \alpha_H^2 \beta_K k^2-4 \alpha_H \beta_K k^2-2 \beta_K k^2-3 a^2 h_{\rm dS}^2 ((\alpha_B+1)^2-\beta_K)^2) \nonumber \\
&-\frac{1}{\alpha_B^2+2 \alpha_B-\beta_K+1} (8 (\alpha_B+1) (9 a^2 (2 (\alpha_H+1) (\alpha_B^2+2 \alpha_B-\beta_K+1) (f_1 (\alpha_B+1) \nonumber \\
&-6 f_2 h_{\rm dS} (2 \alpha_B^2+3 \alpha_B-2 \beta_K+1)) k^2+(9 a^2 (f_1+6 f_2 h_{\rm dS}) (\alpha_B^2+2 \alpha_B-\beta_K+1) h_{\rm dS}^2 \nonumber \\
&+k^2 (\alpha_B-\alpha_H) (2 f_2 h_{\rm dS} (7 \alpha_H+\alpha_B (6 \alpha_H+2)-9)-f_1 (\alpha_H+3))) ((\alpha_B+1)^2 \nonumber \\
&-\beta_K)) ((\alpha_B+1)^2-\beta_K) h_{\rm dS}^2+6 k^2 (\alpha_B+1)^2 (-3 a^2 (2 f_2 h_{\rm dS} (2 \alpha_B^2+(6 \alpha_H+7) \alpha_B+18 \alpha_H-28) \nonumber \\
&-f_1 (\alpha_B+10)) (\alpha_B^2+2 \alpha_B-\beta_K+1) h_{\rm dS}^2-k^2 (\alpha_B-\alpha_H) (2 f_2 h_{\rm dS} ((6 \alpha_B+7) \alpha_H^2+(10 \alpha_B-1) \alpha_H \nonumber \\
&-4 (\alpha_B+1) \beta_H)-f_1 \alpha_H (\alpha_H+3)))+3 k^2 (\alpha_B+1)^2 (\alpha_H (-3 a^2 (2 f_2 h_{\rm dS} (\alpha_B+\alpha_H-8) \nonumber \\
&+f_1 (\alpha_B+\alpha_H-2)) (\alpha_B^2+2 \alpha_B-\beta_K+1) h_{\rm dS}^2-k^2 (\alpha_B-\alpha_H) \alpha_H (f_1 (\alpha_H-1) \nonumber \\
& +2 f_2 h_{\rm dS} (8 \alpha_B+\alpha_H+5)))-3 a^2 h_{\rm dS}^2 (f_1 \alpha_H (\alpha_B+2 \alpha_H-2)+2 f_2 h_{\rm dS} (2 \alpha_B^2+(25 \alpha_H+4) \alpha_B \nonumber \\
&+2 (\alpha_H^2+8 \alpha_H+1))) (\alpha_B^2+2 \alpha_B-\beta_K+1))-6 (3 a^2 (f_1+6 f_2 h_{\rm dS}) (6 a^2 h_{\rm dS}^2 (\alpha_B^2+2 \alpha_B-\beta_K+1) \nonumber \\
&-k^2 (\alpha_B+1) (\alpha_H+1)) ((\alpha_B+1)^2-\beta_K) h_{\rm dS}^2+k^2 (2 f_2 h_{\rm dS} (7 \alpha_H+\alpha_B (6 \alpha_H+2)-9) \nonumber \\
&-f_1 (\alpha_H+3)) ((\alpha_H+1) (6 a^2 h_{\rm dS}^2 (\alpha_B^2+2 \alpha_B-\beta_K+1)-k^2 (\alpha_B+1) (\alpha_H+1)) \nonumber \\
&-3 a^2 h_{\rm dS}^2 (\alpha_B-2 \alpha_H-1) ((\alpha_B+1)^2-\beta_K))) (\alpha_B^2+2 \alpha_B-\beta_K+1)+k^2 (\alpha_B+1)^2 \nonumber \\
& \times (-27 a^2 (\alpha_B^2+2 \alpha_B-\beta_K+1) (3 f_1-2 f_2 h_{\rm dS} (10 \alpha_H+\alpha_B (6 \alpha_H+2)+2 \beta_K-9)) h_{\rm dS}^2 \nonumber \\
& +3 k^2 (\alpha_B-\alpha_H) \alpha_H (3 f_1+2 f_2 h_{\rm dS} (-12 \alpha_B-2 \beta_B+4 \beta_H+3))+k^2 (\alpha_B-\alpha_H) (15 f_1 \nonumber \\
&-2 f_2 h_{\rm dS} (20 \alpha_B+40 \alpha_H+2 \beta_B-12 \beta_H-45))))))
\end{align}

\begin{align}
a^2& C_3 = -\frac{1}{48  k^5 (\alpha_B+1)^6 a} ((\lambda (c-t)^2+12 m^2) (c-t)^2 (\alpha_B+1) (k^2 (\alpha_B+1)^2 \alpha_H^3+3 k^2 (\alpha_B+1)^2 \alpha_H^2 \nonumber \\
&+3 k^2 (\alpha_B+1)^2 \alpha_H-9 a^2 h_{\rm dS}^2 ((\alpha_B+1)^2-\beta_K)^2 \alpha_H+k^2 (\alpha_B+1)^2-9 a^2 h_{\rm dS}^2 ((\alpha_B+1)^2-\beta_K)^2) a^2 \nonumber \\
&+48 f_2 k^2 (\alpha_B+1) (-12 k^2 (\alpha_B+1)^2 \alpha_H^3-k^2 (\alpha_B+1)^2 (4 \alpha_H+3) \alpha_H^2+2 k^2 (\alpha_B+1)^2 \nonumber \\
&\times (3 \alpha_H^2-4 \alpha_H+4 \beta_H) \alpha_H-9 a^2 h_{\rm dS}^2 (8 \alpha_H^2-4 \alpha_H+5) ((\alpha_B+1)^2-\beta_K)^2 \nonumber \\
&+k^2 (\alpha_B+1)^2 (12 \alpha_H^3+42 \alpha_H^2+(8-24 \beta_H) \alpha_H-4 \beta_H+3))+24 (27 h_{\rm dS}^3 (f_1 \nonumber + 2 f_2 h_{\rm dS} \\
&\times  (2 \alpha_B+4 \alpha_H+3)) ((\alpha_B+1)^2-\beta_K)^3 a^4+18 f_2 h_{\rm dS}^2 (3-2 \alpha_H) (6 a^2 h_{\rm dS}^2 (\alpha_B^2+2 \alpha_B-\beta_K+1) \nonumber \\
&-k^2 (\alpha_B+1) (\alpha_H+1)) ((\alpha_B+1)^2-\beta_K)^2 a^2-3 h_{\rm dS} k^2 (\alpha_B+1)^2 (f_1 (-\alpha_H^2+2 \alpha_H+3) \nonumber \\
&+2 f_2 h_{\rm dS} (105 \alpha_H^2+4 \alpha_H+\alpha_B (24 \alpha_H-2)+2 \beta_B-12 \beta_H+17)) (\alpha_B^2+2 \alpha_B-\beta_K+1) a^2 \nonumber \\
&-2 f_2 k^2 (\alpha_B+1)^2 (\alpha_H^2-12 \alpha_H+4 \beta_H+3) (6 a^2 h_{\rm dS}^2 (\alpha_B^2+2 \alpha_B-\beta_K+1)-k^2 (\alpha_B+1) (\alpha_H+1))) \nonumber \\
&-\frac{1}{1-3 \alpha_H}(8 a^2 (-\alpha_B-1) (27 a^2 (3 \alpha_H-1) ((\alpha_B+1)^2-\beta_K)^2 (6 f_2 h_{\rm dS} (2 \alpha_B^2+3 \alpha_B-2 \beta_K+1) \nonumber \\
&-f_1 (\alpha_B+1)) h_{\rm dS}^3-9 k^2 (\alpha_B+1)^2 \alpha_H (3 \alpha_H-1) (f_1 \alpha_H (\alpha_B+2 \alpha_H-2)+2 f_2 h_{\rm dS} (2 \alpha_B^2 \nonumber \\
&+(25 \alpha_H+4) \alpha_B+2 (\alpha_H^2+8 \alpha_H+1))) h_{\rm dS}-9 k^2 (\alpha_B+1)^2 (1-3 \alpha_H) (f_1 \alpha_H (2 \alpha_B+3 \alpha_H+20) \nonumber \\
&+2 f_2 h_{\rm dS} ((2-8 \alpha_H) \alpha_B^2+(-36 \alpha_H^2-70 \alpha_H+24 \beta_H+4) \alpha_B-51 \alpha_H^2+4 \alpha_H+24 \beta_H+2)) h_{\rm dS} \nonumber \\
&-18 (1-3 \alpha_H) (2 f_2 h_{\rm dS} (7 \alpha_H+\alpha_B (6 \alpha_H+2)-9)-f_1 (\alpha_H+3)) (6 a^2 h_{\rm dS}^2 (\alpha_B^2+2 \alpha_B-\beta_K+1) \nonumber \\
&-k^2 (\alpha_B+1) (\alpha_H+1)) ((\alpha_B+1)^2-\beta_K) h_{\rm dS}-9 k^2 (\alpha_B+1)^2 (3 \alpha_H-1) (2 f_2 h_{\rm dS} (12 \alpha_B^2 \nonumber \\
&+(-18 \alpha_H^2+30 \alpha_H+2 \beta_B-4 \beta_H+29) \alpha_B-2 (15 \alpha_H^2+(-3 \beta_B+6 \beta_H+3 \beta_K-29) \alpha_H-2 \beta_B \nonumber \\
&+8 \beta_H+24))-3 f_1 (\alpha_B+6)) h_{\rm dS}+k^2 (\alpha_B+1)^2 (3 \alpha_H-1) (96 f_2^{(3)} h_{\rm dS}^2+3 (-15 f_1+6 f_2 h_{\rm dS} \nonumber \\
&\times (18 \alpha_H \alpha_B+6 \alpha_B+30 \alpha_H+4 \beta_B-16 \beta_H+6 \beta_K-17)+8 f_1^{(3)}) h_{\rm dS}+8 f_0^{(3)}))))
\end{align}

\acknowledgments
We wish to thank Marco Crisostomi, Vincent Vennin and Filippo Vernizzi for comments on an earlier version of this work. 
We acknowledge the use of the \textsc{xPand} package \footnote{http://www.xact.es/xPand/} \cite{Pitrou:2013hga} for computing the perturbations. 
AL acknowledges funding by the LabEx ENS-ICFP: ANR-10-LABX-0010/ANR-10-IDEX-0001-02 PSL*. This project has received funding /support from the European Union’s Horizon 2020 research and innovation programme under the Marie Skłodowska-Curie grant agreement No 860881-HIDDeN.

\bibliography{Bibliography}{}

\end{document}